\documentclass[aps,prd,twocolumn,superscriptaddress]{revtex4}
\usepackage{epsfig,epsf}
\usepackage{amsmath}
\usepackage{amsthm}
\usepackage{amsfonts}
\usepackage{amssymb}
\usepackage{dsfont}
\usepackage{multirow}
\usepackage{appendix}
\usepackage{slashed}
\usepackage[active]{srcltx}
\usepackage{psfrag}

\begin{document}

\title{Decays of the vector charmonium and bottomonium hybrids }
\date{\today}
\author{B. Barsbay\textsuperscript}
\thanks{Corresponding Author: bbarsbay@dogus.edu.tr}
\affiliation{Division of Optometry, School of Medical Services and Techniques, Dogus
University, Dudullu-\"{U}mraniye, 34775 Istanbul, T\"{u}rkiye}

\begin{abstract}
The full widths of the vector charmonium and bottomonium hybrid mesons $H_{
\mathrm{c}}$ and $H_{\mathrm{b}}$, characterized by the quantum numbers $1^{
\mathrm{--}}$, are determined by analyzing their dominant strong decay
modes: $H_{\mathrm{c}} \to D^{+}D^{-}$, $D_{0}\overline{D}_{0}$, $
D_{s}^{+}D_{s}^{-} $, $ D^{\ast +}D^{\ast -} $, $%
D^{\ast 0}\overline{D}^{\ast 0}$, $ D^{\ast +}D^{-} $, $ D^{\ast 0}\overline{D}^{0} $, $D_{s}^{\ast +}D_{s}^{-}  $  and $H_{\mathrm{b}} \to B^{+}B^{-}$, $B_{0}\overline{B}
_{0}$. To evaluate the partial widths of these channels, we
employ the QCD three-point sum rule approach, which provides a reliable
method for extracting the strong coupling constants at the relevant
hybrid-meson-meson interaction vertices. Based on this analysis, the full
widths of these hybrid quarkonia are found to be $\Gamma _{H_{\mathrm{c}}} =(309.6\pm 39.0)~
\mathrm{MeV} $ and $\Gamma _{H_{\mathrm{b}}} =(78.8\pm 15.4)~\mathrm{MeV} $
. These results are expected to facilitate the interpretation of future
experimental data concerning the spectroscopy and decay patterns of exotic
charmonium- and bottomonium-like hybrid mesons.
\end{abstract}

\maketitle

%%%%%%%%%%%%%%%%%%%%%%%%%%%%%%%%%%%%%%%%%%%%%%%%%%%%%%%%%%%%%%%%%

\section{Introduction}

\label{sec:Intro}
%%%%%%%%%%%%%%%%%%%%%%%%%%%%%%%%%%%%%%%%%%%%%%%%%%%%%%%%%%%

Over the past five decades, Quantum Chromodynamics (QCD) has emerged as the
fundamental theory describing the strong interaction, one of the four
fundamental forces in nature \cite%
{tHooft:1972tcz,Politzer:1973fx,Gross:1973id,Fritzsch:1973pi,Gross:2022hyw}.
QCD has provided profound insights into the structure of hadrons and laid
the theoretical foundation for hadron spectroscopy. Conventional hadrons,
classified as mesons (composed of a quark and an antiquark) and baryons
(composed of three quarks) have been successfully described within this
framework. However, our understanding remains incomplete, particularly
regarding the role of gluon dynamics in the non-perturbative, low-energy
regime of QCD. In this context, the gluon field is expected to play a more
significant role than merely mediating the strong force. To address this
limitation, recent research has increasingly focused on exotic
configurations such as hybrid hadrons, in which gluons act as explicit,
dynamic constituents of the bound state. Investigating such hybrid states
not only challenges and extends the traditional quark model but also opens
promising avenues for uncovering the rich and complex structure of hadronic
matter.

Among the resonances identified in experimental studies, only a few are
regarded as viable candidates for hybrid mesons. These mesons, which possess
unconventional quantum numbers, challenge existing theoretical approaches
and offer a unique opportunity to explore the role of gluonic excitations
within hadronic structures. In particular, the resonances with quantum
numbers $J^{\mathrm{PC}}=$ ${1^{-+}}$, including $\pi _{1}(1600)$ \cite%
{E852:1998mbq}, $\pi _{1}(2015)$ \cite{E852:2004gpn}, and the recently
observed $\eta _{1}(1855)$ \cite{BESIII:2022riz,BESIII:2022iwi}, have
attracted significant attention due to their potential hybrid nature.

The long-standing ambiguity surrounding the nature of the $\pi _{1}(1400)$
and $\pi _{1} (1600)$ resonances has been substantially clarified through
advanced coupled-channel analyses, indicating that experimental data can be
adequately described by considering only the $\pi _{1}(1600)$ state \cite%
{JPAC:2018zyd,Kopf:2020yoa}. This development represents a significant step
forward in understanding mesonic states and their relationship to underlying
quark-gluon dynamics. Furthermore, experimental evidence supporting the
existence of $\pi _{1}(2015)$ has been reported in Refs. \cite%
{E852:2004gpn,BESIII:2022iwi}, providing further insights into the spectrum
of hybrid mesons. Lattice QCD calculations of radially excited states have
also identified $\pi _{1}(2015)$ as a promising candidate for the first
excited state of a hybrid meson, suggesting that it plays a crucial role in
the development of a more comprehensive hybrid meson model \cite%
{Dudek:2010wm}. In addition to these discoveries, the $\eta _{1}(1855)$,
observed through partial-wave analysis of the radiative decay $J/\psi
\rightarrow \gamma \eta _{1}(1855)\rightarrow \gamma \eta \eta ^{\prime }$
\cite{BESIII:2022riz,BESIII:2022iwi}, represents the first isoscalar
particle to be observed with quantum numbers $J^{\mathrm{PC}}=$ ${1^{-+}}$.
This finding is particularly significant, as it provides a new platform for
studying exotic hadronic states and the role of gluonic excitations in
hadron formation. The identification of $\eta _{1} (1855)$ has stimulated
extensive theoretical investigations aimed at elucidating its properties,
internal structure, and broader implications for non-perturbative QCD
phenomena ( see, Refs.\ \cite%
{Chen:2022qpd,Shastry:2022mhk,Wan:2022xkx,Dong:2022cuw,Yan:2023vbh,Yang:2022rck,Qiu:2022ktc}%
).

A number of heavy resonances observed experimentally have been proposed as
potential candidates for hybrid mesons. Notably, the $\psi (4230)$ and $\psi
(4360)$ resonances have been suggested to correspond either to vector hybrid
charmonium states $\overline{c}gc$ or to mesons with substantial exotic
hybrid components \cite{Kou:2005gt,Olsen:2017bmm}. A detailed compilation of
additional resonances that are likely hybrid quarkonia can be found in Ref.\
\cite{Brambilla:2022hhi}.

The hybrid quarkonia $\overline{b}gb$, $\overline{c}gc$ and the hyrid mesons
$\overline{b} gc$ have been extensively studied within various theoretical
frameworks \cite%
{Govaerts:1985fx,Govaerts:1984hc,Zhu:1998sv,Page:1998gz,HadronSpectrum:2012gic,Harnett:2012gs,Qiao:2010zh,Chen:2013zia,Chen:2013eha,Cheung:2016bym,Palameta:2018yce,Miyamoto:2019oin,Brambilla:2018pyn,Woss:2020ayi,TarrusCastella:2021pld,Ryan:2020iog,Soto:2023lbh,Bruschini:2023tmm,Wang:2025ypo}%
. These analyses focus on essential properties of heavy hybrid systems,
including the determination of their spectroscopic parameters, investigation
of decay channels, and characterization of production mechanisms in
different interaction regimes. The employed methodologies include various
quark-gluon models, lattice QCD computations, and QCD sum rules.

The spectroscopic parameters of the scalar, pseudoscalar, vector and
axial-vector hybrid bottomonia $\overline{b}gb$, charmonia $\overline{c}gc$
and mesons $\overline{b}gc$ were also investigated in the framework of the
QCD sum rule method\ \cite{Alaakol:2024zyh}. Furthermore, in Refs.\ \cite%
{Agaev:2025llz,Agaev:2025mqe}, the tensor charmonia $\overline{c}gc$ with $%
J^{\mathrm{PC}}=2^{-+}$ and $2^{++}$ and tensor hybrid mesons $\overline{b}%
gc $ with $J^{\mathrm{P}}=$ ${2^{-}}$ and ${2^{+}}$ were examined, in which
their masses and decay widths were computed.

In the present work, the full widths of the vector charmonium and
bottomonium hybrid mesons $H_{\mathrm{c}}$ and $H_{\mathrm{b}}$,
characterized by quantum numbers $J^{\mathrm{PC}}=$ ${1^{--}}$, are computed
through the analysis of their kinematically allowed decay channels. The
results indicate that $H_{\mathrm{c}}$ primarily decays into conventional
mesons via the processes $H_{\mathrm{V}}\rightarrow D^{+}D^{-}$, $D_{0}%
\overline{D}_{0}$, $D_{s}^{+}D_{s}^{-}$, $ D^{\ast +}D^{\ast -} $, $%
D^{\ast 0}\overline{D}^{\ast 0}$, $ D^{\ast +}D^{-} $, $ D^{\ast 0}\overline{D}^{0} $, $D_{s}^{\ast +}D_{s}^{-}$, while $H_{\mathrm{b}}$ decays
through the channels $H_{\mathrm{b}}\rightarrow B^{+}B^{-}$, $B_{0}\overline{%
B}_{0}$. The partial decay widths of these channels are determined using the
QCD three-point SR method. This approach is crucial for extracting the
strong coupling constants at the hybrid-meson-meson vertices, thereby
allowing for a reliable calculation of the decay widths for the processes
under investigation.

This work is structured in the following manner: In Secs.\ \ref{sec:Hcdecays}%
-\ref{sec:Hcdecays2}, we explore the decay channels of the vector charmonium hybrid meson 
$H_{%
\mathrm{c}}$ and compute partial widths of the processes $H_{\mathrm{c}} \to D^{+}D^{-}$, $D_{0}\overline{D}_{0}$, $
D_{s}^{+}D_{s}^{-} $, $ D^{\ast +}D^{\ast -} $, $%
D^{\ast 0}\overline{D}^{\ast 0}$, $ D^{\ast +}D^{-} $, $ D^{\ast 0}\overline{D}^{0} $, and $D_{s}^{\ast +}D_{s}^{-}  $. The full
width of $H_{\mathrm{c}}$ is also determined in these sections. A similar
analysis for the bottomonium hybrid meson $H_{\mathrm{b}}$ is presented in
Sec.\ \ref{sec:Hbdecays3}, where we evaluate the contributions of the decays $%
H_{\mathrm{b}} \to B^{+}B^{-}$, and $B_{0} \overline{B}_{0}$ to full width
of $H_{\mathrm{b}}$. The last Sec.\ \ref{sec:Dis} contains our concluding
notes.

%%%%%%%%%%%%%%%%%%%%%%%%%%%%%%%%%%%%%%%%%%%%%%%%%%%%%%%%%%%%%%%%%%%%

\section{Decays $H_{\mathrm{c}} \to D^{+}D^{-}$, $D_{0}\overline{D}_{0}$,
and $D_{s}^{+}D_{s}^{-}$}

\label{sec:Hcdecays}

%%%%%%%%%%%%%%%%%%%%%%%%%%%%%%%%%%%%%%%%%%%%%%%%%%%%%%%%%%%

In this section, we calculate the widths of the decays $H_{\mathrm{c}%
}\rightarrow D^{+}D^{-}$, $D_{0}\overline{D}_{0}$, and $D_{s}^{+}D_{s}^{-}$,
where $D$ mesons are pseudoscalar particles. The partial widths of these
processes are determined by the strong coupling constants $g_{l}(\ l=1-3)$,
which describe the interactions between the hybrid meson $H_{\mathrm{c}}$
and the final-state mesons at the relevant three-particle vertices.
Accordingly, the central focus of this section is the evaluation of these
couplings.

\begin{widetext}

\begin{figure}[h!]
\begin{center}

\includegraphics[totalheight=6cm,width=8cm]{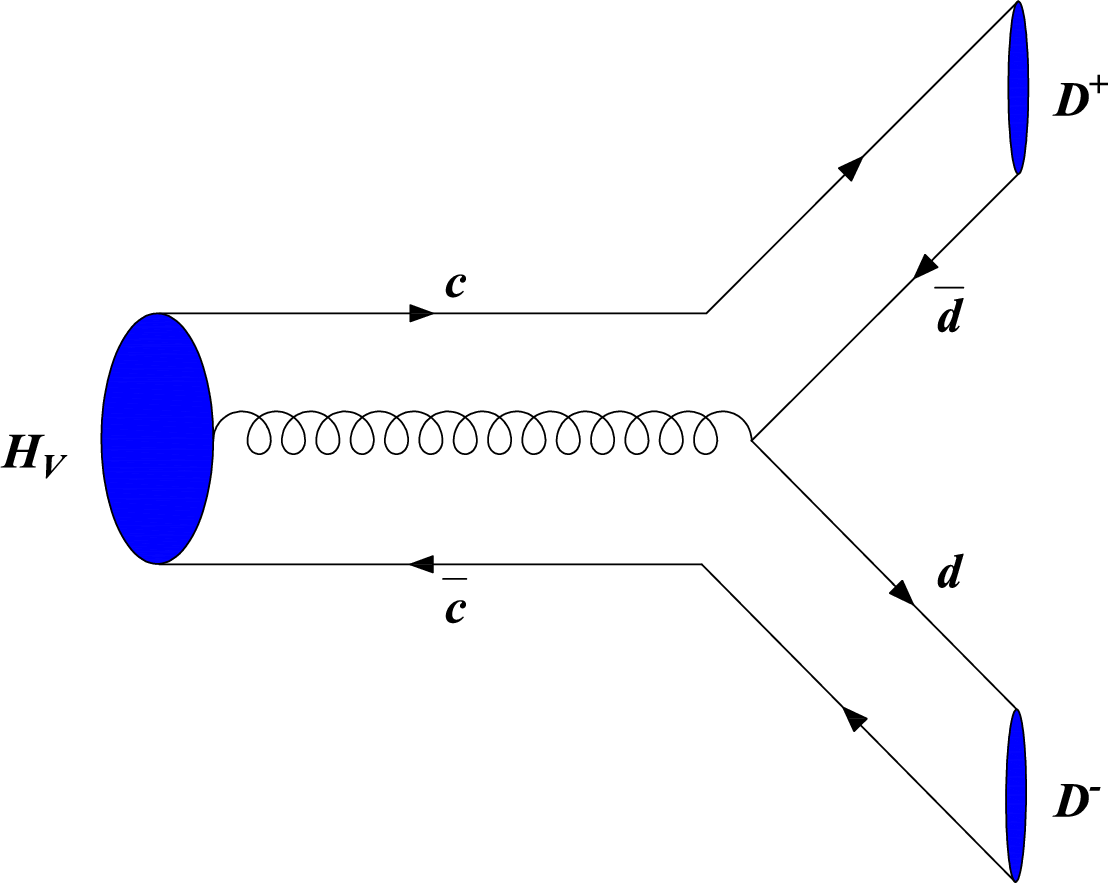}

\vspace{0.6cm} % 

\includegraphics[totalheight=6cm,width=8cm]{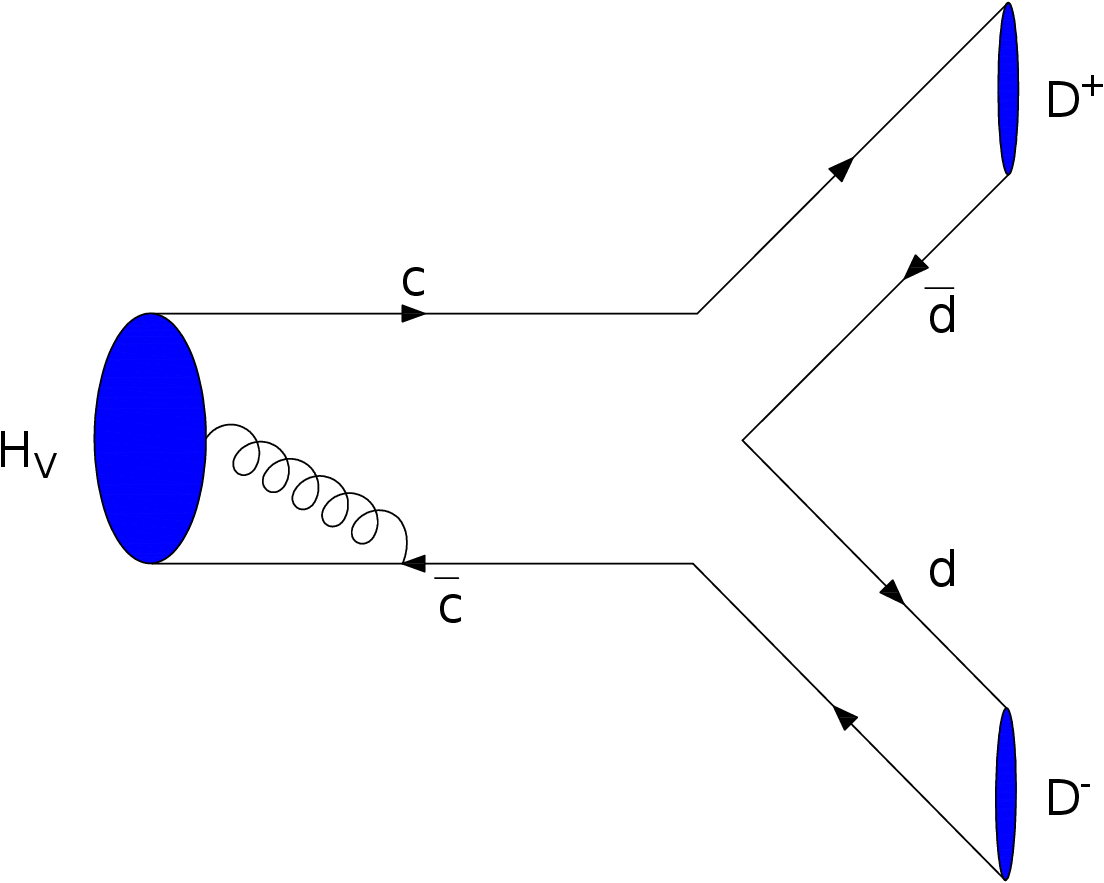}%
\hspace{0.5cm}%
\includegraphics[totalheight=6cm,width=8cm]{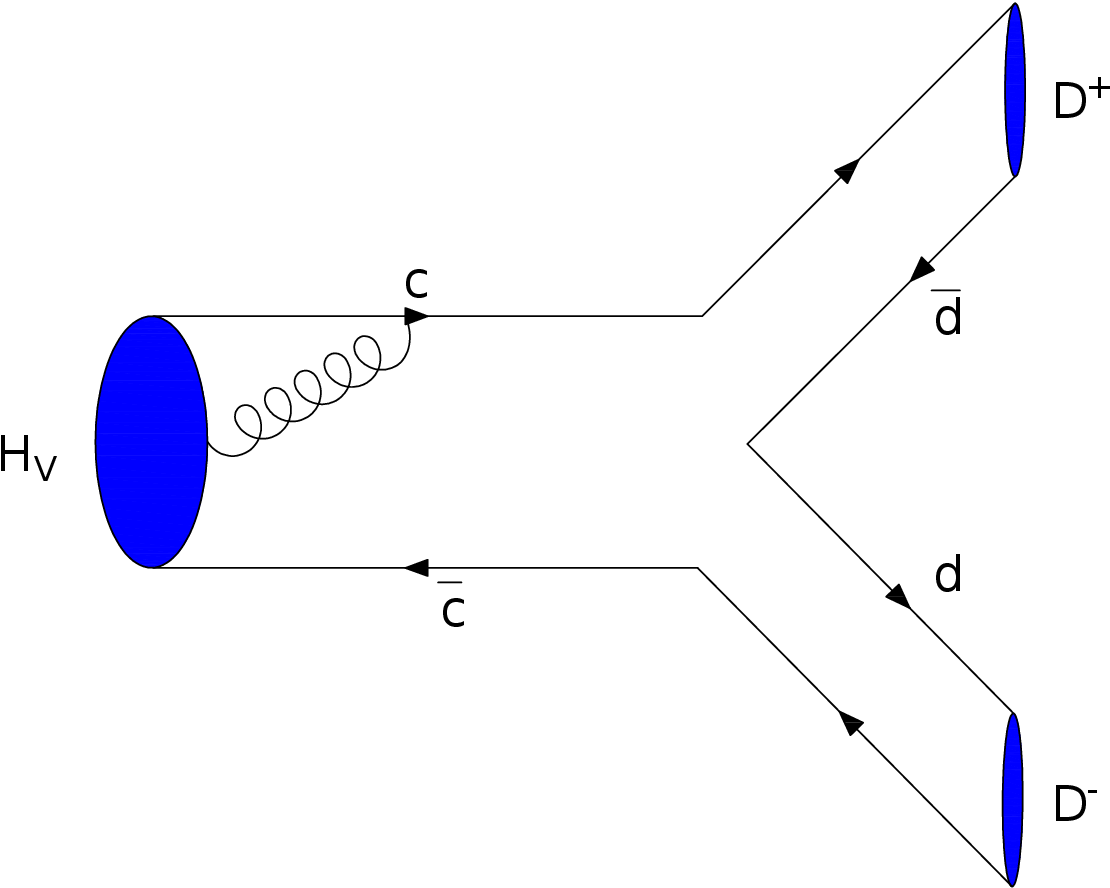}

\end{center}
\caption{Sample perturbative diagrams corresponding to the strong decay $H_{\mathrm{c}%
}\rightarrow D^{+}D^{-}$. }
\label{fig:decay}
\end{figure}

\end{widetext}

In the decay process $H_{\mathrm{c}}\rightarrow D^{+}D^{-}$, the strong
coupling constant $g_{1}$ plays a central role, as it is depicted in Fig.\ %
\ref{fig:decay}. In the subsequent analysis, we focus in detail on this
particular channel, while for the other decay modes, we limit ourselves to
presenting the essential formulas and numerical outcomes.

The strong coupling $g_{1}$ can be obtained from the three-point correlation
function%
\begin{eqnarray}
\Pi _{\mu }(p,p^{\prime }) &=&i^{2}\int d^{4}xd^{4}ye^{ip^{\prime
}x}e^{iqy}\langle 0|\mathcal{T}\{J^{D^{+}}(x)  \notag \\
&&\times J^{D^{-}}(y)J_{\mu }^{\dagger }(0)\}|0\rangle ,  \label{eq:CF1}
\end{eqnarray}%
where $J_{\mu }(x)\ $ is the interpolating current for the vector charmonium
hybrid meson $H_{\mathrm{c}}$
\begin{equation}
J_{\mu }(x)=g_{s}\overline{c}_{a}(x)\gamma ^{\theta }\gamma _{5}\frac{{%
\lambda }_{ab}^{n}}{2}\widetilde{G}_{\mu \theta }^{n}(x)c_{b}(x),
\label{eq:CRHc}
\end{equation}%
In Eq.\ (\ref{eq:CRHc}), $g_{s}$ denotes the QCD strong coupling constant,
and $c_{a}(x)$ represents the $c$ quark field. The indices $a$ and $b$ label
color degrees of freedom, while ${\lambda }^{n}$, $n=1,2,..8$ are the
Gell-Mann matrices. The dual field of the gluon field strength tensor is
shown by $\widetilde{G}_{\mu \theta }^{n}(x)=\varepsilon _{\mu \theta \alpha
\beta }G^{n\alpha \beta }(x)/2$.

The expressions for the currents $J^{D^{+}}(x)$ and $J^{D^{-}}(x)$,
corresponding to the $D^{+}$ and $D^{-}$ mesons, are given as follows:
\begin{equation}
J^{D^{+}}(x)=\overline{d}_{j}(x)i\gamma _{5}c_{j}(x),\ J^{D^{-}}(x)=%
\overline{c}_{i}(x)i\gamma _{5}d_{i}(x),  \label{eq:CRD}
\end{equation}%
where $i,\ j=1,2,3$ are color indices.

Following the sum rule framework, the function $\Pi _{\mu }(p,p^{\prime })$
have to be expressed in terms of the parameters of the participating
particles. This allows us to extract the physical side of the sum rule. For
this purpose, we present $\Pi _{\mu }(p,p^{\prime })$ in the form presented
below
\begin{eqnarray}
&&\Pi _{\mu }^{\mathrm{Phys}}(p,p^{\prime })=\frac{\langle
0|J^{D^{+}}|D^{+}(p^{\prime })\rangle }{p^{\prime 2}-m_{D}^{2}}\frac{\langle
0|J^{D^{-}}|D^{-}(q)\rangle }{q^{2}-m_{D}^{2}}  \notag \\
&&\times \langle D^{-}(q)D^{+}(p^{\prime })|H_{\mathrm{c}}(p,\varepsilon )\rangle \frac{%
\langle H_{\mathrm{c}}(p,\varepsilon )|J_{\mu }^{\dagger }|0\rangle }{p^{2}-m_{H_{\mathrm{c}%
}}^{2}}+\cdots .  \label{eq:CF2}
\end{eqnarray}%
In this case, the explicit contribution comes solely from ground-state
particles, with the effects of higher resonances and continuum states being
depicted by ellipses. It is evident the four-momenta of $H_{\mathrm{c}}$ and
$D^{+}$ particles are represented by $p$ and $p^{\prime }$, respectively.
Hence, the momentum of the $D^{-}$ meson amounts to $q=p-p^{\prime }$.

In order to simplify the correlation function $\Pi _{\mu }^{\mathrm{Phys}%
}(p,p^{\prime })$, we rewrite the matrix elements that appear in Eq.\ (\ref%
{eq:CF1}) by expressing them in terms of the masses and decay constants of
the participating particles. Specifically, for the vector hybrid meson $H_{%
\mathrm{c}}$, the matrix element $\langle 0|J_{\mu }|H_{\mathrm{c}%
}(p,\varepsilon )\rangle $ can be substituted by the product of its mass $%
m_{H_{\mathrm{c}}}$ and current coupling $f_{H_{\mathrm{c}}}$
\begin{equation}
\langle 0|J_{\mu }|H_{\mathrm{c}}(p,\varepsilon )\rangle =m_{H_{\mathrm{c}%
}}f_{H_{\mathrm{c}}}\varepsilon _{\mu },  \label{eq:ME1}
\end{equation}%
where $\varepsilon _{\mu }$ is the polarization vector of $H_{\mathrm{c}}$.

The matrix element of the pseudoscalar $D$ mesons is given by the relation
\begin{equation}
\langle 0|J^{D}|D\rangle =\frac{f_{D}m_{D}^{2}}{m_{c}},  \label{eq:ME2}
\end{equation}%
with $m_{D}$ and $f_{D}$ corresponding to the mass and decay constant of the
$D$ meson. Here, $m_{c}$ is the $c$ quark mass.

The vertex $\langle D^{+}(p^{\prime })D^{-}(q)|H_{\mathrm{c}}(p,\varepsilon
)\rangle $ has the following form
\begin{equation}
\langle D^{+}(p^{\prime })D^{-}(q)|H_{\mathrm{c}}(p,\varepsilon )\rangle
=g_{1}(q^{2})\varepsilon (p)\cdot p^{\prime }.  \label{eq:ME3}
\end{equation}%
Here, $g_{1}(q^{2})$ denotes the form factor, which evaluates the strong
coupling $g_{1}$ when the transferred momentum squared matches the mass
shell condition of the $D^{-}$ meson, i.e., at $q^{2}=m_{D}^{2}$.

Taking these expressions into account, one can easily transform $\Pi _{\mu
}^{\mathrm{Phys}}(p,p^{\prime })$ into the following expression:
\begin{eqnarray}
&&\Pi _{\mu }^{\mathrm{Phys}}(p,p^{\prime })=g_{1}(q^{2})\frac{f_{H_{\mathrm{
c}}}m_{H_{\mathrm{c}}}f_{D}^{2}m_{D}^{4}}{m_{c}^{2}\left( p^{2}-m_{H_{
\mathrm{c}}}^{2}\right) (p^{\prime 2}-m_{D}^{2})}  \notag \\
&&\times \frac{1}{(q^{2}-m_{D}^{2})}\left[ \frac{(m_{H_{\mathrm{c}
}}^{2}+m_{D}^{2}-q^{2})}{2m_{H_{\mathrm{c}}}^{2}}p_{\mu }-p_{\mu }^{\prime }
\right] +\cdots .  \label{eq:PhysSide1}
\end{eqnarray}
where the dots denote contributions of higher resonances and continuum
states. As is seen, the correlator $\Pi _{\mu }^{\mathrm{Phys}}(p,p^{\prime
})$ contains two Lorentz structures $p_{\mu }$ and $p_{\mu }^{\prime }$ .
One of these structures can be chosen to proceed with the sum rule analysis.
To extract the sum rule for $g_{1}(q^{2})$, we work with the term
proportional to $p_{\mu }$, and represent the corresponding invariant
amplitude by $\Pi _{\mu }^{\mathrm{Phys}}(p^{2},p^{\prime 2},q^{2})$.

The second component in the derivation of the sum rule for $g_{1}(q^{2})$ is
the evaluation of the correlation function Eq.\ (\ref{eq:CF1}), which should
be computed using the quark/gluon propagators. The correlation function
within the operator product expansion (OPE) framework takes the form
\begin{eqnarray}
&&\Pi _{\mu }^{\mathrm{OPE}}(p,p^{\prime })=\frac{\epsilon _{\mu \theta
\alpha \beta }}{2}\int d^{4}xd^{4}ye^{ip^{\prime }x}e^{iqy}g_{s}\frac{{
\lambda }_{ab}^{n}}{2}G_{\alpha \beta }^{n}(0)  \notag \\
&&\times \mathrm{Tr}\left[ S_{c}^{ai}(-y)\gamma _{5}S_{d}^{ij}(y-x)\gamma
_{5}S_{c}^{jb}(x)\gamma _{5}\gamma _{\theta }\right] ,  \label{eq:QCDside1}
\end{eqnarray}
where $S_{c(d)}(x)$ are $d$ and $c$ quark propagators
\begin{eqnarray}
&&S_{d}^{ab}(x)=i\delta _{ab}\frac{\slashed x}{2\pi ^{2}x^{4}}-\delta _{ab}
\frac{m_{d}}{4\pi ^{2}x^{2}}-\delta _{ab}\frac{\langle \overline{d}d\rangle
}{12}  \notag \\
&&+i\delta _{ab}m_{d}\frac{\slashed x\langle \overline{d}d\rangle }{48}
-\delta _{ab}\frac{x^{2}}{192}\langle \overline{d}g_{s}\sigma Gd\rangle
\notag \\
&&+i\delta _{ab}m_{d}\frac{x^{2}\slashed x}{1152}\langle \overline{d}
g_{s}\sigma Gd\rangle -i\frac{g_{s}G_{ab}^{\alpha ^{\prime }\beta ^{\prime }}
}{32\pi ^{2}x^{2}}\left[ \slashed x{\sigma _{\alpha ^{\prime }\beta ^{\prime
}}+\sigma _{\alpha ^{\prime }\beta ^{\prime }}}\slashed x\right]   \notag \\
&&-i\delta _{ab}\frac{x^{2}\slashed xg_{s}^{2}\langle \overline{d}d\rangle
^{2}}{7776}-\delta _{ab}\frac{x^{4}\langle \overline{d}d\rangle \langle
g_{s}^{2}G^{2}\rangle }{27648}+\cdots ,  \label{eq:pPro}
\end{eqnarray}
and
\begin{eqnarray}
&&S_{c}^{ab}(x)=i\int \frac{d^{4}k}{(2\pi )^{4}}e^{-ikx}\Bigg \{\frac{\delta
_{ab}\left( {\slashed k}+m_{c}\right) }{k^{2}-m_{c}^{2}}  \notag \\
&&-\frac{g_{s}G_{ab}^{\alpha ^{\prime }\beta ^{\prime }}}{4}\frac{\sigma
_{\alpha ^{\prime }\beta ^{\prime }}\left( {\slashed k}+m_{c}\right) +\left(
{\slashed k}+m_{c}\right) \sigma _{\alpha ^{\prime }\beta ^{\prime }}}{
(k^{2}-m_{c}^{2})^{2}}  \notag \\
&&+\frac{g_{s}^{2}G^{2}}{12}\delta _{ab}m_{c}\frac{k^{2}+m_{c}{\slashed k}}{
(k^{2}-m_{c}^{2})^{4}}+\cdots \Bigg \}.  \label{eq:QProp}
\end{eqnarray}
Above, we have adopted the short-hand notations
\begin{equation}
G_{ab}^{\alpha ^{\prime }\beta ^{\prime }}\equiv G_{m}^{\alpha ^{\prime
}\beta ^{\prime }}\lambda _{ab}^{m}/2,\ \ G^{2}=G_{\alpha ^{\prime }\beta
^{\prime }}^{m}G_{m}^{\alpha ^{\prime }\beta ^{\prime }}.\
\end{equation}

\begin{widetext}

\begin{figure}[h!]
\begin{center}

\includegraphics[totalheight=6cm,width=8cm]{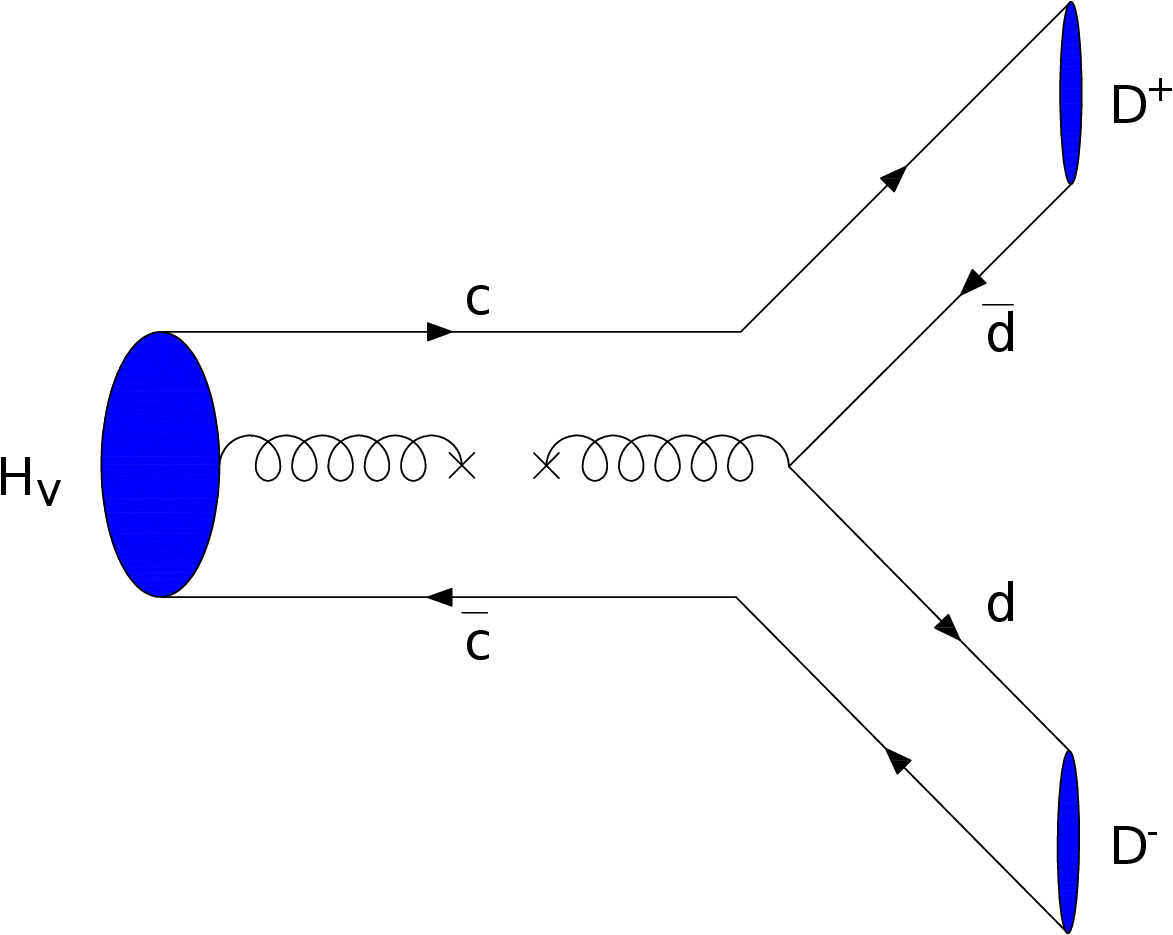}

\vspace{0.6cm}

\includegraphics[totalheight=6cm,width=8cm]{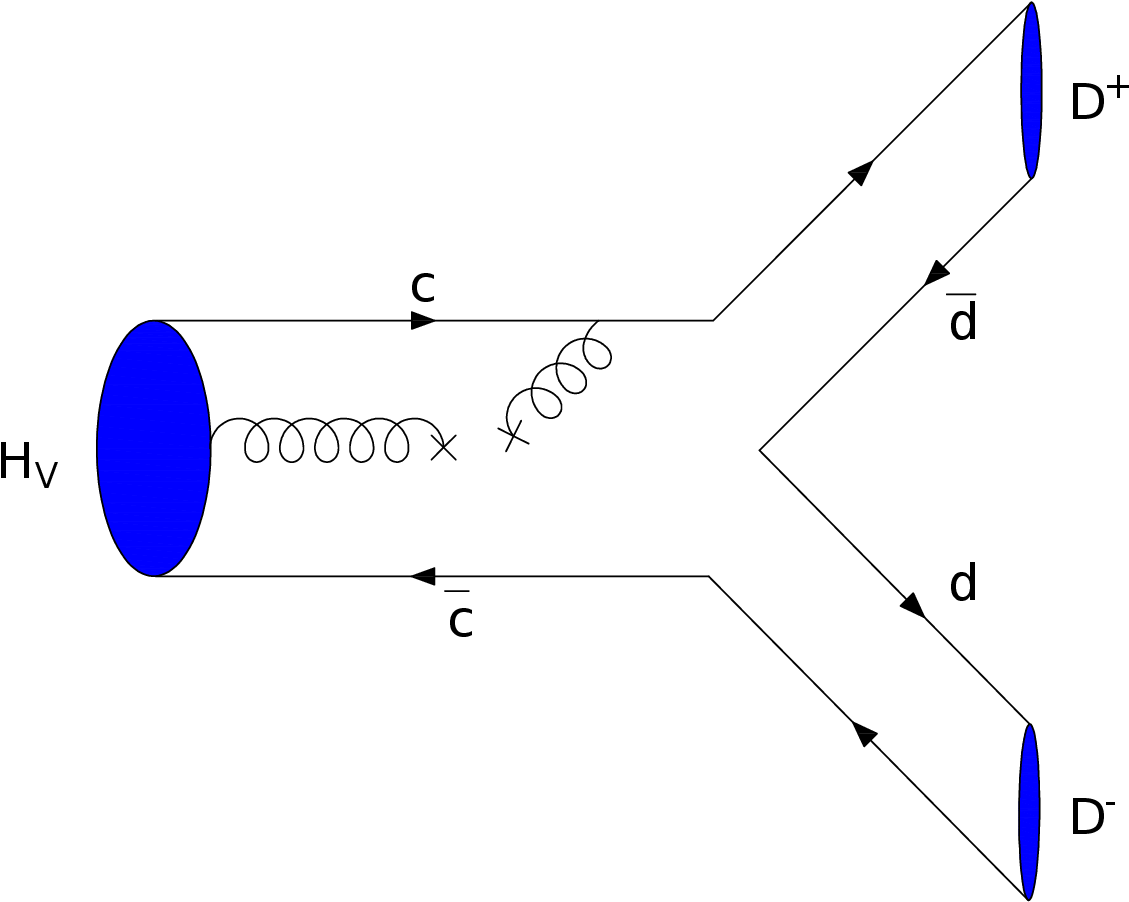}
\hspace{0.5cm}
\includegraphics[totalheight=6cm,width=8cm]{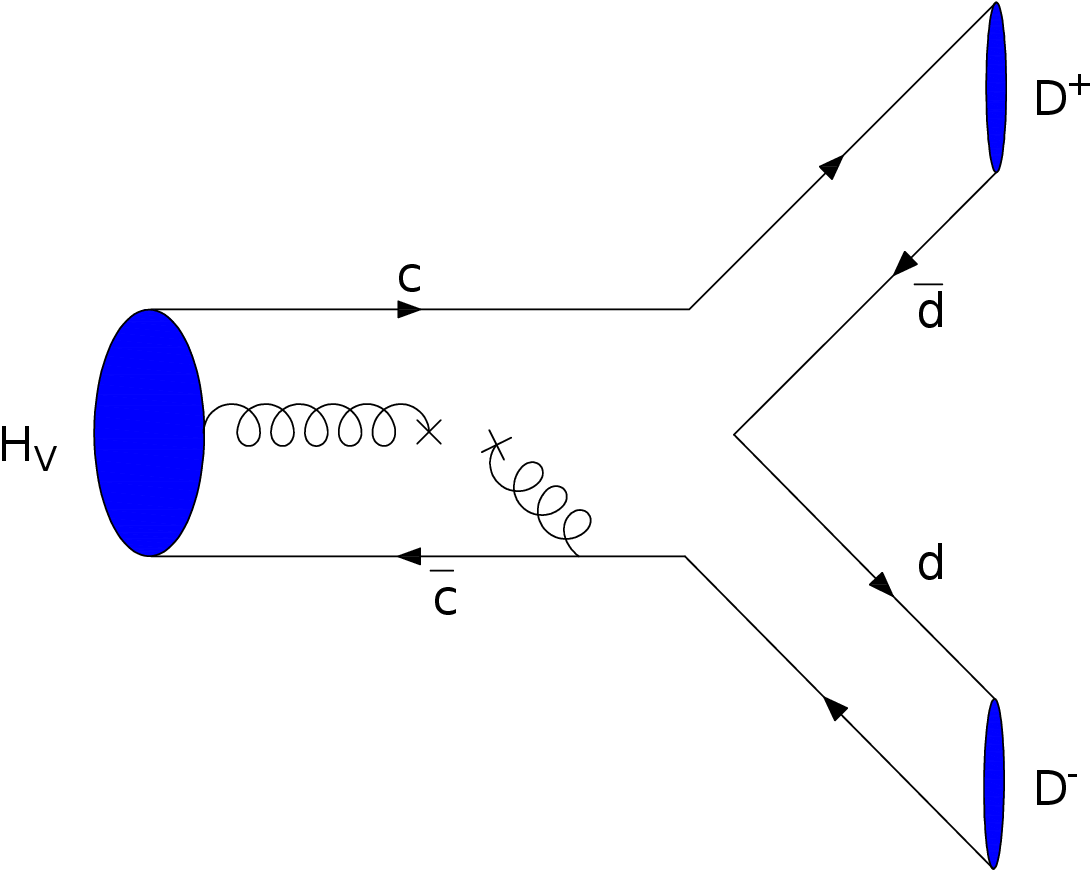}

\end{center}
\caption{Sample diagrams of the two-gluon condensate contributions.}
\label{fig:GGcond}
\end{figure}

\end{widetext}

In Eq.\ (\ref{eq:QCDside1}), $\Pi _{\mu }^{\mathrm{OPE}}(p,p^{\prime })$
includes three quark propagators and the gluon field strength tensor $%
G_{\alpha \beta }^{n}(0)$. When this gluon tensor contracts with one of the
terms $-\frac{ig_{s}G_{ab}^{\alpha ^{\prime }\beta ^{\prime }}}{32\pi
^{2}x^{2}}\left[ \slashed x\sigma _{\alpha ^{\prime }\beta ^{\prime
}}+\sigma _{\alpha ^{\prime }\beta ^{\prime }}\slashed x\right] $ or $-\frac{%
g_{s}G_{ab}^{\alpha ^{\prime }\beta ^{\prime }}}{4}\frac{\sigma _{\alpha
^{\prime }\beta ^{\prime }}\left( {\ \slashed k}+m_{c}\right) +\left( {%
\slashed k}+m_{c}\right) \sigma _{\alpha ^{\prime }\beta ^{\prime }}}{%
(k^{2}-m_{Q}^{2})^{2}}$ from the quark propagators, it generates the matrix
element of two-gluon fields sandwiched between the vacuum states $\langle
0|G_{\alpha ^{\prime }\beta ^{\prime }}^{m}(x)G_{\alpha \beta
}^{n}(0)|0\rangle $. This two- gluon matrix element is analyzed using two
distinct approaches. Initially, it is treated as the full gluon propagator
in coordinate space connecting points $0$ and $x$ (see Fig.\ 
\ref{fig:decay}), applying the relation
\begin{eqnarray}
&&\langle 0|G_{\alpha ^{\prime }\beta ^{\prime }}^{m}(x)G_{\alpha \beta
}^{n}(0)|0\rangle =\frac{\delta ^{mn}}{2\pi ^{2}x^{4}}\left[ g_{\beta
^{\prime }\beta }\left( g_{\alpha ^{\prime }\alpha }-\frac{4x_{\alpha
^{\prime }}x_{\alpha }}{x^{2}}\right) \right.   \notag \\
&&+(\beta ^{\prime },\beta )\leftrightarrow (\alpha ^{\prime },\alpha
)-\beta ^{\prime }\leftrightarrow \alpha ^{\prime }-\beta \leftrightarrow
\alpha ].
\end{eqnarray}%
Alternatively, this matrix element $\langle 0|G_{\alpha ^{\prime }\beta
^{\prime }}^{m}(x)G_{\alpha \beta }^{n}(0)|0\rangle $ is interpreted as the
two-gluon condensate. In this approach, one expands the gluon field at point
$x$ around $x=0$ and retains only the leading term. Consequently, we get
\begin{eqnarray}
&&\langle 0|g_{s}^{2}G_{\alpha ^{\prime }\beta ^{\prime }}^{m}(x)G_{\alpha
\beta }^{n}(0)|0\rangle =\frac{\langle g_{s}^{2}G^{2}\rangle }{96}\delta
^{mn}[g_{\alpha ^{\prime }\alpha }g_{\beta ^{\prime }\beta }  \notag
\label{eq:Gcond} \\
&&-g_{\alpha ^{\prime }\beta }g_{\alpha \beta ^{\prime }}].
\end{eqnarray}%
The two-gluon condensate diagrams contributing to the correlation function are shown 
in Fig.\ 
\ref{fig:GGcond}.

Following these steps, the resulting expressions are combined with the
remaining two quark propagators and other relevant factors to complete the
calculation. We denote by $\Pi ^{\mathrm{OPE}}(p^{2},p^{\prime 2},q^{2})$
the invariant amplitude in $\Pi _{\mu }^{\mathrm{OPE}}(p,p^{\prime })$ that
corresponds to the structure proportional to $p_{\mu }$.

To obtain sum rule for the form factor $g_{1}(q^{2})$, we equate the
invariant amplitudes $\Pi ^{\mathrm{Phys}}(p^{2},p^{\prime 2},q^{2})$ and $%
\Pi ^{\mathrm{OPE}}(p^{2},p^{\prime 2},q^{2})$, thereby establishing the
corresponding sum rule relation. The contributions from higher resonances
and the continuum can be effectively suppressed by performing Borel
transformations with respect to the variables $-p^{2}$ and $-p^{\prime 2}$
on both sides of the equation. These undesired terms are then subtracted
under the quark-hadron duality assumption. After carrying out these steps,
we arrive at the final expression:
\begin{eqnarray}
&&g_{1}(q^{2})=\frac{2m_{H_{\mathrm{c}}}m_{c}^{2}(q^{2}-m_{D}^{2})}{f_{H_{%
\mathrm{c}}}f_{D}^{2}m_{D}^{4}(m_{H_{\mathrm{c}}}^{2}+m_{D}^{2}-q^{2})}
\notag \\
&&\times e^{m_{H_{\mathrm{c}}}^{2}/M_{1}^{2}}e^{m_{D}^{2}/M_{2}^{2}}\Pi (%
\mathbf{M}^{2},\mathbf{s}_{0},q^{2}).  \label{eq:SRCoup}
\end{eqnarray}%
Here, $\Pi (\mathbf{M}^{2},\mathbf{s}_{0},q^{2})$ represents the QCD side of
the correlation function $\Pi ^{\mathrm{OPE}}(p^{2},p^{\prime 2},q^{2})$,
evaluated after the application of Borel transformations and the subtraction
of continuum contributions. This quantity can be formulated 
\begin{eqnarray}
&&\Pi (\mathbf{M}^{2},\mathbf{s}_{0},q^{2})=\int_{4m_{c}^{2}}^{s_{0}}ds%
\int_{m_{c}^{2}}^{s_{0}^{\prime }}ds^{\prime }\rho (s,s^{\prime },q^{2})
\notag \\
&&\times e^{-s/M_{1}^{2}}e^{-s^{\prime }/M_{2}^{2}}+\Pi (\mathbf{M}^{2}),  \label{eq:SCoupl}
\end{eqnarray}%
where $(M_{1}^{2},s_{0})$ and $(M_{2}^{2},s_{0}^{\prime })$ correspond to
the Borel and continuum subtraction parameters for the $H_{\mathrm{c}}$ and $%
D^{+}$ channels, respectively. It should also be emphasized that the
spectral density $\rho (s,s^{\prime },q^{2})$ is obtained by evaluating the
imaginary part of the invariant amplitude $\Pi ^{\mathrm{OPE}
}(p^{2},p^{\prime 2},q^{2})$ with respect to the variables $p^{2}$ and $%
p^{\prime 2}$. The second component of the invariant amplitude $\Pi (\mathbf{M}^{2})$ 
contains nonperturbative contributions extracted directly from $ \Pi ^{\mathrm{OPE}}
(p^{2},p^{\prime 2},q^{2}) $ through double Borel transformations. As an example, the explicit expression of $\Pi(\mathbf{M}^{2},\mathbf{s}_{0},q^{2})$ for the perturbative part and for the nonperturbative parts of dimensions 3 and 4 is presented in the Appendix. In our analysis, however, nonperturbative terms are taken into account up to dimension 8. Also as examples, Figs.~\ref{fig:decay} and \ref{fig:GGcond} present illustrative diagrams corresponding to the perturbative part and the dimension-4 contribution, respectively.

It is evident that the form factor $g_{1}(q^{2})$ explicitly depends on the
mass $m_{H_{\mathrm{c}}}$ and current coupling  $f_{H_{\mathrm{c}}}$ of the
hybrid meson $H_{\mathrm{c}}$, both of which were determined in Ref.\ \cite%
{Alaakol:2024zyh}.
\begin{eqnarray}
m_{H_{\mathrm{c}}} &=&(4.12\pm 0.11)~\mathrm{GeV},  \notag \\
f_{H_{\mathrm{c}}} &=&(4.0\pm 0.4)\times 10^{-2}~\mathrm{GeV}^{3}.
\label{eq:Result1}
\end{eqnarray}%
For this analysis, the two-point sum rule formalism was employed, wherein
the Borel and continuum subtraction parameters were constrained to the
following intervals
\begin{equation}
M^{2}\in \lbrack 4,4.6]~\mathrm{GeV}^{2},\ s_{0}\in \lbrack 24,26]~\mathrm{%
GeV}^{2}.  \label{eq:Wind1}
\end{equation}

The sum rule Eq.\ (\ref{eq:SRCoup}) also depends on the mass $%
m_{D}=(1869.5\pm 0.05)~\mathrm{MeV}$ and decay constant $f_{D}=(211.9\pm
1.1)~\mathrm{MeV}$ of the $D^{\pm }$ mesons \cite{Agaev:2024wvp}. The gluon
condensate and $c$ quark mass are well known parameters
\begin{eqnarray}
&&\langle \frac{\alpha _{s}G^{2}}{\pi }\rangle =(0.012\pm 0.004)~\mathrm{GeV}%
^{4},  \notag \\
&&m_{c}=(1.27\pm 0.02)~\mathrm{GeV}.  \label{Parameters}
\end{eqnarray}

To carry out the numerical analysis, it is necessary to determine
appropriate working intervals for the parameters $(M_{1}^{2},s_{0})$ and $%
(M_{2}^{2},s_{0}^{\prime })$. For the pair $(M_{1}^{2},s_{0})$ associated
with the hybrid meson $H_{\mathrm{c}}$, we adopt the ranges presented in
Eq.\ (\ref{eq:Wind1}). It should be emphasized that the intervals specified
in Eq.\ (\ref{eq:Wind1}) fully satisfy all the constraints dictated by the
sum rule formalism.

To ensure the reliability of the sum rule results in the $D^{+}$ channel,
the parameters $(M_{2}^{2},\ s_{0}^{\prime })$ are chosen within limits
\begin{equation}
M_{2}^{2}\in \lbrack 1.5,3]~\mathrm{GeV}^{2},\ s_{0}^{\prime }\in \lbrack
5,5.2]~\mathrm{GeV}^{2}.  \label{eq:Wind2}
\end{equation}

In order to compute the partial decay width of the process $H_{\mathrm{c}%
}\rightarrow D^{+}D^{-}$, it is essential to determine the form factor $%
g_{1}(q^{2})$ at the mass shell of the $D^{-}$ meson $q^{2}=m_{D}^{2}$. To
achieve this, a fit function $\mathcal{F}_{1}(Q^{2})$ is introduced where $%
Q^{2}=-q^{2}$. This function has accurately to reproduce the sum rule
predictions in the region $Q^{2}>0$, and can be analytically continued to $%
Q^{2}<0$ to evaluate $\mathcal{F}_{1}(-m_{D}^{2})$.

\begin{figure}[h]
\includegraphics[width=8.5cm]{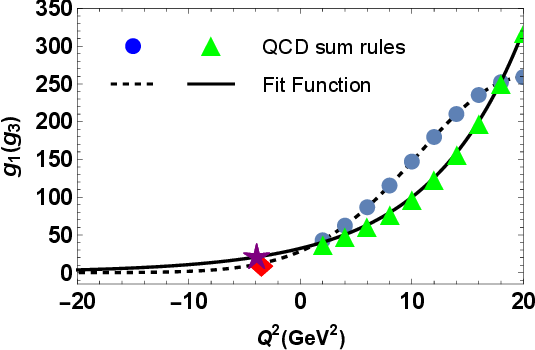}
\caption{SR data and functions $\mathcal{F}_{1}(Q^{2})$ (dashed line) and $%
\mathcal{F}_{3}(Q^{2})$ (solid line). The labels are fixed at the points $%
Q^{2}=-m_{D}^{2}$ and $Q^{2}=-m_{D_{s}}^{2}$. }
\label{fig:Fit1}
\end{figure}

In present article, we use the functions $\mathcal{F}_{l}(Q^{2})$
\begin{equation}
\mathcal{F}_{l}(Q^{2})=\mathcal{F}_{l}^{0}\mathrm{\exp }\left[ c_{l}^{1}%
\frac{Q^{2}}{m^{2}}+c_{l}^{2}\left( \frac{Q^{2}}{m^{2}}\right) ^{2}\right] ,
\label{eq:FitF}
\end{equation}%
with parameters $\mathcal{F}_{l}^{0}$, $c_{l}^{1}$ and $c_{l}^{2}$. To fix
their values, comparison between SR predictions and $\mathcal{F}_{1}(Q^{2})$
is required. The analysis of the form factor $g_{1}(q^{2})$ gives results $%
\mathcal{F}_{1}^{0}=28.31$, $c_{1}^{1}=3.72$, and $c_{1}^{2}=-1.56$. This
function is depicted in Fig.\ \ref{fig:Fit1}, where one can be convinced in
nice agreement of $\mathcal{F}_{1}(Q^{2})$ and QCD SR results.

The value obtained for the coupling $g_{1}$ is
\begin{equation}
g_{1}\equiv \mathcal{F}_{1}(-m_{D}^{2})=12.32\pm 1.60.  \label{eq:Coupl1}
\end{equation}%
The decay width for the process $H_{\mathrm{c}}\rightarrow D^{+}D^{-}$ is
calculated using the following expression:%
\begin{equation}
\Gamma \left[ H_{\mathrm{c}}\rightarrow D^{+}D^{-}\right] =g_{1}^{2}\frac{%
\lambda }{96\pi }\left( 1-\frac{4m_{D}^{2}}{m_{H_{\mathrm{c}}}^{2}}\right) ,
\label{eq:PDw2}
\end{equation}%
where $\lambda =\lambda (m_{H_{\mathrm{c}}},m_{D},m_{D})$, and
\begin{equation}
\lambda (x,y,z)=\frac{\sqrt{%
x^{4}+y^{4}+z^{4}-2(x^{2}y^{2}+x^{2}z^{2}+y^{2}z^{2})}}{2x}.
\end{equation}%
We find
\begin{equation}
\Gamma _{1}\left[ H_{\mathrm{c}}\rightarrow D^{+}D^{-}\right] =(76.8\pm
21.3)~\mathrm{MeV}.  \label{eq:DW1}
\end{equation}

The strong coupling for the decay $H_{\mathrm{c}}\rightarrow D_{0}\overline{D%
}_{0}$ is nearly the same as that for the $H_{\mathrm{c}}\rightarrow
D^{+}D^{-}$ process, with only a small difference in the meson masses. The
mass of the $D_{0}$ meson is $m_{D_{0}}=(1864.84\pm 0.05)~\mathrm{MeV}$,
which slightly differs from the mass of the $D^{\pm }$ mesons. As a result,
the strong coupling $g_{2}(q^{2})$ is approximately equal to $g_{1}(q^{2})$.
Accordingly, the partial decay width of the process $H_{\mathrm{c}%
}\rightarrow D_{0}\overline{D}_{0}$ is found to be
\begin{equation}
\Gamma _{2}\left[ H_{\mathrm{c}}\rightarrow D_{0}\overline{D}_{0}\right]
=(80.3\pm 22.1)~\mathrm{MeV}.  \label{eq:DW2}
\end{equation}

To analyze the $H_{\mathrm{c}}\rightarrow D_{s}^{+}D_{s}^{-}$ process, some
technical modifications are necessary. The first step involves specifying
the interpolating currents of the mesons $D_{s}^{+}$ and $D_{s}^{-}$, which
are defined as
\begin{equation}
J^{D_{s}^{+}}(x)=\overline{s}_{j}(x)i\gamma _{5}c_{j}(x),\ J^{D_{s}^{-}}(x)=%
\overline{c}_{i}(x)i\gamma _{5}s_{i}(x).
\label{eq:CRD2}
\end{equation}%
The matrix element corresponding to the $D_{s}^{+}$ and $D_{s}^{-}$ mesons
is given by
\begin{equation}
\langle 0|J^{D_{s}}|D_{s}\rangle =\frac{f_{D_{s}}m_{D_{s}}^{2}}{m_{c}+m_{s}},
\label{eq:ME4}
\end{equation}%
where the strange quark mass is taken as $m_{s}=(93.5\pm 0.8)~\mathrm{MeV}$.
In this expression, the parameters $m_{D_{s}}=(1969.0\pm 1.4)~\mathrm{MeV}$
and $f_{D_{s}}=(249.9\pm 0.5)~\mathrm{MeV}$ denote the mass and decay
constant of the $D_{s} $ meson, respectively \cite{Agaev:2025llz}. As a
result, Eqs.\ (\ref{eq:SRCoup}) and (\ref{eq:SCoupl}) change accordingly,
where one should replace $m_{c}^{2}\rightarrow (m_{c}+m_{s})^{2} $.

In the evaluation of the form factor $g_{3}(q^{2})$, we adopt specific
choices for the Borel masses and continuum thresholds. For the $H_{\mathrm{c}%
}$ channel, we employ the parameters $(M_{1}^{2},s_{0})$ as defined in Eq.\ (%
\ref{eq:Wind1}). For the $D_{s}^{+}$ channel, the working regions are taken
to be
\begin{equation}
M_{2}^{2}\in \lbrack 2.5,3.5]~\mathrm{GeV}^{2},\ s_{0}^{\prime }\in \lbrack
5,6]~\mathrm{GeV}^{2},  \label{eq:Wind3}
\end{equation}%
To facilitate the extraction of $g_{3}$, we use a fit function $\mathcal{F}%
_{3}(Q^{2})$, characterized by the parameters $\mathcal{F}_{3}^{0}=32.21$, $%
c_{3}^{1}=1.89$, and $c_{3}^{2}=0.05$ (see, Fig.\ \ref{fig:Fit1}). Utilizing
this fit, the value of the strong coupling is obtained as
\begin{equation}
g_{3}\equiv \mathcal{F}_{3}(-m_{D_{s}}^{2})=20.96\pm 2.73,  \label{eq:Coupl3}
\end{equation}%
and the corresponding partial width for the decay $H_{\mathrm{c}}\rightarrow
D_{s}^{+}D_{s}^{-}$ is calculated to be
\begin{equation}
\Gamma _{3}\left[ H_{\mathrm{c}}\rightarrow D_{s}^{+}D_{s}^{-}\right]
=(77.0\pm 21.3)~\mathrm{MeV}.  \label{eq:DW3}
\end{equation}

The decay modes analyzed in this section provide a basis for estimating the
full decay width of the vector hybrid charmonium $H_{\mathrm{c}}$, which is
found to be
\begin{equation}
\Gamma _{H_{\mathrm{c}}}=(234.1\pm 37.4)~\mathrm{MeV}.  \label{eq:fullDW1}
\end{equation}

%%%%%%%%%%%%%%%%%%%%%%%%%%%%%%%%%%%%%%%%%%%%%%%%%%%%%%%%%%%%%%%%%%%%

\section{ Processes $H_{\mathrm{c}}\rightarrow D^{\ast +}D^{\ast -}$, and  $%
D^{\ast 0}\overline{D}^{\ast 0}$}

\label{sec:Hcdecays1}

%%%%%%%%%%%%%%%%%%%%%%%%%%%%%%%%%%%%%%%%%%%%%%%%%%%%%%%%%%%

Here, we perform a detailed analysis of the decay $H_{\mathrm{c}}\rightarrow 
D^{\ast+}D^{\ast -}$. For this channel, the sum rule for the strong form factor 
$g_{4}(q^{2})$ at the vertex $H_{\mathrm{c}%
}D^{\ast +}D^{\ast -}$ is derived from the corresponding correlation function,
\begin{eqnarray}
\Pi _{\mu \nu \mu{\prime } }(p,p^{\prime }) &=&i^{2}\int
d^{4}xd^{4}ye^{ip^{\prime }y}e^{iqx}\langle 0|\mathcal{T}\{J_{\mu }^{D^{\ast
+}}(x)  \notag \\
&&\times J_{\nu }^{D^{\ast - }}(y)J_{\mu{\prime } }^{\dagger }(0)\}|0\rangle ,  \label{eq:CFstr}
\end{eqnarray}%
where
\begin{equation}
J_{\mu }^{D^{\ast +}}(x)=\overline{d}_{j}(x)\gamma _{\mu }c_{j}(x),\ J_{\nu }^{D^{\ast -}}(x)=\overline{c}_{i}(x)\gamma _{\nu }d_{i}(x),  \label{eq:CRB}
\end{equation}%
are interpolating currents for the vector mesons $D^{\ast +}$ and $D^{\ast -}$, respectively.

The correlation function $\Pi _{\mu \nu \mu{\prime }}(p,p^{\prime })$ can be expressed in terms of the physical parameters characterizing the particles participating in this decay process
\begin{eqnarray}
&&\Pi _{\mu \nu \mu{\prime } }^{\mathrm{Phys}}(p,p^{\prime })=%
\frac{\langle 0|J_{\mu }^{D^{\ast +}}|D^{\ast +}(p^{\prime },\varepsilon
)\rangle }{p^{\prime 2}-m_{D^{\ast }}^{2}}\frac{\langle 0|J_{\nu
}^{D^{\ast }}|D^{\ast -}(q,\varepsilon )\rangle }{q^{2}-m_{D^{\ast }}^{2}%
}  \notag \\
&&\times \langle D^{\ast +}(p^{\prime },\varepsilon )D^{\ast
-}(q,\varepsilon )|H_{\mathrm{c}}(p,\epsilon )\rangle \frac{%
\langle H_{\mathrm{c}}(p,\epsilon )|J_{\mu{\prime }}^{\dagger }|0\rangle }{p^{2}-m_{H_{%
\mathrm{c}}}^{2}}+\cdots ,  \notag \\
&&
\end{eqnarray}%
where $\varepsilon _{\mu }(p^{\prime })$ and $\varepsilon _{\nu }(q)$ are the polarization
vectors of the $\ D^{\ast +}$ and $D^{\ast -}$ mesons, respectively.
The matrix elements employed in this part are given by
\begin{eqnarray}
\langle 0|J_{\mu }^{D^{\ast }+}|D^{\ast +}(p^{\prime },\varepsilon
)\rangle &=&f_{D^{\ast }}m_{D^{\ast }}\varepsilon _{\mu }(p^{\prime }),
\notag \\
\langle 0|J_{\nu }^{D^{\ast }-}|D^{\ast -}(q,\varepsilon )\rangle
&=&f_{D^{\ast }}m_{D^{\ast }}\varepsilon _{\nu }(q),
\end{eqnarray}%
where $m_{D^{\ast }}=(2010.26\pm 0.05)\ \mathrm{MeV}$ and $f_{D^{\ast
}}=(223.5\pm 8.4)\ \mathrm{MeV}$ are the mass and decay constant of the
mesons $D^{\ast \pm }$, respectively \cite{Agaev:2025llz}. The vertex $\langle D^{\ast +}(p^{\prime
},\varepsilon )D^{\ast -}(q,\varepsilon )|H_{\mathrm{c}%
}(p,\epsilon )\rangle $ has the following form:
\begin{eqnarray}
&&\langle D^{\ast +}(p^{\prime },\varepsilon )D^{\ast -}(q,\varepsilon
)|H_{\mathrm{c}}(p,\epsilon )\rangle =g_{4}(q^{2})
\notag \\
&&\times \lbrack (q-p^{\prime })_{\gamma }g_{\alpha \beta }-(p+q)_{\alpha
}g_{\gamma \beta }+(p+q)_{\beta }g_{\gamma \alpha }]  \notag \\
&&\times \epsilon ^{\gamma }(p)\varepsilon ^{\ast \alpha }(p^{\prime
})\varepsilon ^{\ast \beta }(q).  \notag \\
&&  \label{eq:HcVVVertex}
\end{eqnarray}

As a result, the physical correlation function $\Pi _{\mu \nu \alpha \beta }^{\mathrm{Phys}%
}(p,p^{\prime })$ is represented by the following comprehensive expression:
\begin{eqnarray}
&&\Pi _{\mu \nu \mu{\prime } }^{\mathrm{Phys}}(p,p^{\prime })=\frac{g_{4}(q^{2})f_{H_{%
\mathrm{c}}}m_{H_{%
\mathrm{c}}}f_{D^{\ast }}^{2}m_{D^{\ast }}^{2}}{(p^{2}-m_{H_{
\mathrm{c}}}^{2})(p^{\prime
2}-m_{D^{\ast }}^{2})(q^{2}-m_{D^{\ast }}^{2})}  \notag \\
&&\times \left[ \frac{m_{H_{%
\mathrm{c}}}^{2}+m_{D^{\ast }}^{2}-q^{2}}{2m_{D^{\ast }}^{4}}p_{\mu{\prime }
}p_{\nu }p_{\mu }^{\prime }-\frac{m_{H_{%
\mathrm{c}}}^{2}}{m_{D^{\ast }}^{2}}g_{\mu \mu{\prime }
}p_{\nu }^{\prime }\right.  \notag \\
&&-\frac{m_{H_{%
\mathrm{c}}}^{2}-2m_{D^{\ast }}^{2}}{m_{D^{\ast }}^{4}}\left( p_{\nu }p_{\mu{\prime }
}^{\prime }-p_{\nu }^{\prime }p_{\mu{\prime } }^{\prime }\right) p_{\mu }^{\prime
}+2g_{\mu \nu }p_{\mu{\prime } }^{\prime }+2g_{\mu{\prime } \nu }p_{\mu }  \notag \\
&&\left. -\left( m_{H_{%
\mathrm{c}}}^{2}+m_{D^{\ast }}^{2}-q^{2}\right) \left( \frac{%
g_{\mu \nu }p_{\mu{\prime } }}{m_{H_{%
\mathrm{c}}}^{2}}+\frac{g_{\mu{\prime } \nu }p_{\mu
}^{\prime }}{m_{D^{\ast }}^{2}}\right) ...\right] .  \label{eq:PhysSide2a}
\end{eqnarray}%

The QCD side of the sum rule can be written as
\begin{eqnarray}
&&\Pi _{\mu \nu \mu{\prime } }^{\mathrm{OPE}}(p,p^{\prime
})=i^{2}\frac{\epsilon _{\mu{\prime } \nu{\prime }
\alpha \beta }}{2}\int d^{4}xd^{4}ye^{ip^{\prime }x}e^{iqy}g_{s}\frac{{%
\lambda }_{ab}^{n}}{2}G_{\alpha \beta }^{n}(0)  \notag \\
&&\times \mathrm{Tr}\left[ S_{c}^{ai}(-y)\gamma _{\mu }S_{d}^{ij}(y-x)\gamma
_{\nu}S_{c}^{jb}(x)\gamma _{5}\gamma _{\nu{\prime } }\right] .
\label{eq:QCDside2a}
\end{eqnarray}%
The sum rule for the form factor $g_{4}(q^{2})$ is derived using the
structure $p_{\mu{\prime }
}p_{\nu }p_{\mu }^{\prime }$ in the correlation functions.

In numerical analysis, the parameters $M_{2}^{2}$ and $s_{0}^{\prime }$ in
the $D^{\ast +}$ meson channel are chosen in the form
\begin{equation}
M_{2}^{2}\in \lbrack 2,4]~\mathrm{GeV}^{2},\ s_{0}^{\prime }\in \lbrack
5.5,6.5]~\mathrm{GeV}^{2}.
\label{eq:Wind1Dstr}
\end{equation}%
The strong coupling $g_{4}$ amounts to
\begin{equation}
{g}_{4}\equiv \mathcal{F}_{4}(-m_{D^{\ast }}^{2})=0.07\pm 0.01.  \label{eq:Coupl4}
\end{equation}%
It has been estimated at the mass shell $q^{2}=m_{D^{\ast }}^{2}$ of the $
D^{\ast -}$ meson by employing the interpolating function $\mathcal{F}_{4}(Q^{2})$ . The function $\mathcal{F}_{4}(Q^{2})$  is determined by the parameters $\mathcal{
F}_{4}^{0}=0.067$, $c_{4}^{1}=0.19$, and $c_{4}^{2}=-0.076$.

The width of this decay is
\begin{equation}
\Gamma \left[ H_{\mathrm{c}}\rightarrow D^{\ast +}D^{\ast -}%
\right] =g_{4}^{2}\frac{\widetilde{\lambda }}{24\pi }\left( \frac{1}{4\widetilde{%
\zeta }^{2}}-\frac{4}{\widetilde{\zeta }}-12\widetilde{\zeta }-17\right) ,
\label{eq:DWstr}
\end{equation}%
where $\widetilde{\lambda }=\lambda (m_{H_{
\mathrm{c}}}%
,m_{D^{\ast }},m_{D^{\ast }})$, and $%
\widetilde{\zeta }=m_{D^{\ast }}^{2}/m_{H_{
\mathrm{c}}}^{2}$. Then, we get
\begin{equation}
\Gamma _{4}\left[ H_{\mathrm{c}}\rightarrow D^{\ast +}D^{\ast -}%
\right] =(30.4\pm 7.7)~\mathrm{MeV}.
\end{equation}

The decay width of the process $H_{\mathrm{c}}\rightarrow D^{\ast 0}\overline{D}^{\ast 0}$ 
is determined by Eq.\ (\ref{eq:DWstr}), as the quark structure of the 
$ D^{\ast 0}\overline{D}^{\ast 0} $ pair can be obtained from that of $ D^{\ast +}D^{\ast -} 
$ through the substitution $d\rightarrow u$. The small mass difference between
$ D^{\ast +}D^{\ast -}  $ and $D^{\ast 0}\overline{D}^{\ast 0}$ mesons is disregarded in the 
analysis.

%%%%%%%%%%%%%%%%%%%%%%%%%%%%%%%%%%%%%%%%%%%%%%%%%%%%%%%%%%%%%%%%%%%%

\section{$H_{\mathrm{c}}\rightarrow D^{\ast +}D^{-}$, $%
D^{\ast 0}\overline{D}^{0}$ and $D_{s}^{\ast +}D_{s}^{-}$}

\label{sec:Hcdecays2}

%%%%%%%%%%%%%%%%%%%%%%%%%%%%%%%%%%%%%%%%%%%%%%%%%%%%%%%%%%%

The form factor $g_{5}(q^{2})$, which describes the strong interaction among the particles at the $H_{\mathrm{%
c}}D^{\ast +}D^{-}$ vertex, is extracted from the analysis of the corresponding correlation function
\begin{eqnarray}
\Pi _{\mu \nu  }(p,p^{\prime }) &=&i^{2}\int
d^{4}xd^{4}ye^{ip^{\prime }y}e^{iqx}\langle 0|\mathcal{T}\{J_{\mu }^{D^{\ast
+}}(x)  \notag \\
&&\times J^{D^{-}}(y)J_{\nu }^{\dagger }(0)\}|0\rangle .  \label{eq:CFstra}
\end{eqnarray}%

The correlation function $\Pi _{\mu \nu }(p,p^{\prime })$ can be expressed in terms of the parameters characterizing the particles participating in this decay process
\begin{eqnarray}
&&\Pi _{\mu \nu }^{\mathrm{Phys}}(p,p^{\prime })=%
\frac{\langle 0|J_{\mu }^{D^{\ast +}}|D^{\ast +}(p^{\prime },\varepsilon
)\rangle }{p^{\prime 2}-m_{D^{\ast}}^{2}}\frac{\langle 0|J^{D^{-}}|D^{-}(q )\rangle }{q^{2}-m_{D}^{2}%
}  \notag \\
&&\times \langle D^{\ast +}(p^{\prime },\varepsilon )D^{-}(q)|H_{\mathrm{c}}(p,\epsilon )\rangle \frac{%
\langle H_{\mathrm{c}}(p,\epsilon )|J_{\nu}^{\dagger }|0\rangle }{p^{2}-m_{H_{%
\mathrm{c}}}^{2}}+\cdots .  \notag \\
&&
\end{eqnarray}%
Here, the matrix elements have been presented in the previous sections, and the vertex $\langle D^{\ast +}(p^{\prime
},\varepsilon )D^{-}(q )|H_{\mathrm{c}%
}(p,\epsilon )\rangle $ is modeled by the equation
\begin{eqnarray}
\langle D^{\ast +}(p^{\prime },\varepsilon )D^{-}(q)|H_{\mathrm{c}}(p,\epsilon
)\rangle &=&g_{5}(q^{2})\varepsilon _{\alpha \beta \mu \nu }\epsilon
_{\alpha }\epsilon
_{\beta }^{\ast }p_{\mu}p_{\nu}^{\prime }.  \notag \\
&&
\end{eqnarray}

Then, the correlator becomes equal to
\begin{eqnarray}
&&\Pi _{\mu \nu  }^{\mathrm{Phys}}(p,p^{\prime })=\frac{g_{4}(q^{2})f_{H_{%
\mathrm{c}}}m_{H_{%
\mathrm{c}}}f_{D}m_{D}^{2}f_{D^{\ast }}m_{D^{\ast }}}{m_{c}(p^{2}-m_{H_{
\mathrm{c}}}^{2})(p^{\prime
2}-m_{D^{\ast }}^{2})(q^{2}-m_{D^{\ast }}^{2})}  \notag \\
&&\times \varepsilon _{\alpha \beta \mu \nu }p_{\alpha}p_{\beta}^{\prime }+\cdots  . \label{eq:PhysSide2b}
\end{eqnarray}%

Expressed through the quark-gluon propagators, the correlation function $ \Pi _{\mu \nu  }(p,p^{\prime
}) $ can be written as
\begin{eqnarray}
&&\Pi _{\mu \nu }^{\mathrm{OPE}}(p,p^{\prime
})=-i\frac{\epsilon _{\nu \theta
\alpha \beta }}{2}\int d^{4}xd^{4}ye^{ip^{\prime }x}e^{iqy}g_{s}\frac{{%
\lambda }_{ab}^{n}}{2}G_{\alpha \beta }^{n}(0)  \notag \\
&&\times \mathrm{Tr}\left[ S_{c}^{ai}(-y)\gamma _{\mu }S_{d}^{ij}(y-x)\gamma
_{5}S_{c}^{jb}(x)\gamma _{5}\gamma _{\theta }\right] .
\label{eq:QCDside2b}
\end{eqnarray}%
The sum rule for the form factor $g_{5}(q^{2})$ is obtained by employing the
 $\varepsilon _{\alpha \beta \mu \nu }p_{\alpha}p_{\beta}^{\prime }$ structure in the correlation functions.
 
 In numerical computations, we choose the parameters $(M_{1}^{2},s_{0})$ and $%
(M_{2}^{2},s_{0}^{\prime })$ in the following manner: In the hybrid channels
we use $(M_{1}^{2},s_{0})$ from Eq.\ (\ref{eq:Wind1}) and $ (M_{2}^{2},s_{0}^{\prime }) $ from Eq.\ (\ref{eq:Wind1Dstr}).
The strong coupling $g_{5}$ is determined at the mass shell $q^{2}=m_{D}^{2}$ of the $%
D^{-}$ meson
\begin{equation}
{g}_{5}\equiv \mathcal{F}_{5}(-m_{D}^{2})=(0.25\pm 0.03)~\mathrm{GeV}^{-1}.  \label{eq:Coupl5}
\end{equation}%
 The interpolation function $\mathcal{F}_{5}(Q^{2})$ is determined by the parameters $\mathcal{%
F}_{5}^{0}=0.22\ \mathrm{GeV}^{-1}$, $c_{5}^{1}=-0.50$, and $c_{5}^{2}=0.027$.

The width of the decay $H_{\mathrm{c}}\rightarrow D^{\ast +}D^{-}$ can
be obtained by means of the formula%
\begin{equation}
\Gamma \lbrack H_{\mathrm{c}}\rightarrow D^{\ast +}D^{-}]=g_{5}^{2}%
\frac{\lambda _{1}}{24m_{H_{
\mathrm{c}}}^{2}}|M|^{2},  \label{eq:VVPSw}
\end{equation}%
where $|M|^{2}$ is%
\begin{eqnarray}
&&|M|^{2} = \frac{1}{2} 
\left( m_{H_{\mathrm{c}}}^{2} + m_{D^{\ast}}^{2} - m_{D}^{2} \right)^{2} 
- 2\, m_{H_{\mathrm{c}}}^{2} m_{D^{\ast}}^{2} .
\end{eqnarray}
In Eq.\ (\ref{eq:VVPSw}), we also use  the function $\lambda _{1}=\lambda
(m_{H_{
\mathrm{c}}},m_{D^{\ast }},m_{D})$.
Then, for the partial width of the process under consideration, we find $\ $%
\begin{equation}
\Gamma _{5} \left[ H_{\mathrm{c}}\rightarrow D^{\ast }{}^{+}D^{-}\right]
=(5.9\pm 1.5)~\mathrm{MeV}.  \label{eq:VVPSDW}
\end{equation}%

The decay width of the process $H_{\mathrm{c}}\rightarrow D^{\ast 0}\overline{D}^{0}$ 
is evaluated using Eq.\ (\ref{eq:VVPSw}), noting that the quark composition of the 
$ D^{\ast 0}\overline{D}^{\ast 0} $ final state can be obtained from that of $ D^{\ast +}
D^{-} $ by the replacement $d\rightarrow u$. In the present analysis, the small mass 
difference between the $ D^{\ast +}D^{-}  $ and $D^{\ast 0}\overline{D}^{0}$ channels is 
neglected.

The analysis of the decay $H_{\mathrm{c}}\rightarrow D_{s}^{\ast +}D_{s}^{-}$ does not differ from that presented above. The interpolating current for the $ D_{s}^{-} $ meson was given in Eq.\ (\ref{eq:CRD2}), while for the $ D_{s}^{\ast +} $  meson it is taken as
\begin{equation}
J_{\mu }^{D_{s}^{\ast }}(x)=\overline{s}_{j}(x)\gamma _{\mu }c_{j}(x).
\end{equation}%
The matrix element of $ D_{s}^{\ast +} $ meson is
\begin{eqnarray}
0|J_{\mu }^{D_{s}^{\ast }}|D_{s}^{\ast -}(p^{\prime },\varepsilon )\rangle
&=&f_{D_{s}^{\ast }}m_{D_{s}^{\ast }}\varepsilon _{\mu }(p^{\prime }).
\end{eqnarray}%
Here, $m_{D_{s}^{\ast }}=(2112.2\pm 0.4)~\mathrm{MeV}$ and
$f_{D_{s}^{\ast }}=(268.8\pm 6.5)~\mathrm{MeV}$ are the mass and decay constant of the $ D_{s}^{\ast +} $ meson, respectively \cite {Agaev:2025llz}.

In the numerical analysis, the parameters $M_{1}^{2}$ and $s_{0}$
are fixed to the values given in Eq.\ (\ref{eq:Wind1}). For the $D_{s}^{\ast }$
channel, the working regions are chosen as
\begin{equation}
M_{2}^{2}\in \lbrack 2.5,3.5]~\mathrm{GeV}^{2},\ s_{0}^{\prime }\in \lbrack
6,8]~\mathrm{GeV}^{2}.
\end{equation}%
The sum rule predictions for the form factor $g_{6}(q^2)$ care well described by the extrapolation function $\mathcal{F}%
_{6}(Q^{2})$ with fit parameters $\mathcal{F}_{6}^{0}=0.66\ \mathrm{GeV}^{-1}$, $%
c_{6}^{1}=-0.96$, and $c_{6}^{2}=0.20$. 

As a result, for the strong coupling $g_{6}$, we get
\begin{equation}
{g}_{6}\equiv \mathcal{F}_{6}(-m_{D_{s}}^{2})=(0.81\pm 0.10)~\mathrm{GeV}^{-1}.  \label{eq:Coupl6}
\end{equation}%
Computations yield for the partial width of this decay
\begin{equation}
\Gamma _{6}\left[ H_{\mathrm{c}}\rightarrow D_s^{\ast +}D_s^{ -}%
\right] =(2.9\pm 0.7)~\mathrm{MeV}.
\end{equation}

The partial decay widths presented in the preceding three sections serve to estimate the total width of the vector hybrid charmonium $H_{\mathrm{c}}$.
\begin{equation}
\Gamma \left[ {H}_{\mathrm{c}}\right] =(310.34\pm 39)~\mathrm{MeV}.
\end{equation}

%%%%%%%%%%%%%%%%%%%%%%%%%%%%%%%%%%%%%%%%%%%%%%%%%%%%%%%%%%%%%%%%%%%%

\section{ Channels $H_{\mathrm{b}} \to B^{+}B^{-}$, and $B_{0}\overline{B}%
_{0} $}

\label{sec:Hbdecays3}

%%%%%%%%%%%%%%%%%%%%%%%%%%%%%%%%%%%%%%%%%%%%%%%%%%%%%%%%%%%

The decays of the vector bottomonium hybrid meson $H_{\mathrm{b}}$ to
pseudoscalar mesons $B^{+}B^{-}$, and $B_{0}\overline{B}_{0}$ can be
analyzed within a modified version of the formalism presented in the
previous section. For illustrative purposes, the specific channel $H_{%
\mathrm{b}}\rightarrow B^{+}B^{-}$ is considered, and the corresponding form
factor $g_{7}(q^{2})$ is extracted that characterizes the vertex $H_{\mathrm{%
b}}B^{+}B^{-}$.

The correlation function employed to obtain the sum rule for $g_{7}(q^{2})$
is given by%
\begin{eqnarray}
\Pi _{\mu }(p,p^{\prime }) &=&i^{2}\int d^{4}xd^{4}ye^{ip^{\prime
}x}e^{iqy}\langle 0|\mathcal{T}\{J^{B^{+}}(x)  \notag \\
&&\times J^{B^{-}}(y)J_{\mu }^{\dagger }(0)\}|0\rangle ,  \label{eq:CF2A}
\end{eqnarray}%
where
\begin{equation}
J^{B^{+}}(x)=\overline{d}_{j}(x)i\gamma _{5}b_{j}(x),\ J^{B^{-}}(x)=%
\overline{b}_{i}(x)i\gamma _{5}d_{i}(x),  \label{eq:CRB}
\end{equation}%
are interpolating currents for the pseudoscalar mesons $B^{+}$ and $B^{-}$,
respectively.

To construct the phenomenological side of the sum rule the relevant matrix
elements are defined as
\begin{eqnarray}
0|J_{\mu }|H_{\mathrm{b}}(p,\varepsilon )\rangle  &=&m_{H_{\mathrm{b}}}f_{H_{%
\mathrm{b}}}\varepsilon _{\mu },  \notag \\
\langle 0|J^{B}|B\rangle  &=&\frac{f_{B}m_{B}^{2}}{m_{b}}.
\end{eqnarray}%
Here, $m_{B}=(5279.41\pm 0.07)~\mathrm{MeV}$ and $f_{B}=(206\pm 7)~\mathrm{%
MeV}$ are the mass and decay constants of the $B^{\pm }$ mesons \cite%
{PDG:2022,Narison:2012xy}. The mass of $b$ quark is taken as $%
m_{b}=(4.183\pm 0.007)~\mathrm{GeV}$.

In this case, the correlators $\Pi _{\mu }^{\mathrm{Phys}}(p,p^{\prime })$
and $\Pi _{\mu }^{\mathrm{OPE}}(p,p^{\prime })$ are given by the forms after
evident replacements of the masses, decay constants and quark propagators
\begin{eqnarray}
&&\Pi _{\mu }^{\mathrm{Phys}}(p,p^{\prime })=g_{1}(q^{2})\frac{f_{H_{\mathrm{%
b}}}m_{H_{\mathrm{b}}}f_{B}^{2}m_{B}^{4}}{m_{b}^{2}\left( p^{2}-m_{H_{%
\mathrm{b}}}^{2}\right) (p^{\prime 2}-m_{B}^{2})}  \notag \\
&&\times \frac{1}{(q^{2}-m_{B}^{2})}\left[ \frac{(m_{H_{\mathrm{b}%
}}^{2}+m_{B}^{2}-q^{2})}{2m_{H_{\mathrm{b}}}^{2}}p_{\mu }-p_{\mu }^{\prime }%
\right] +\cdots   \label{eq:PhysSide2}
\end{eqnarray}%
and
\begin{eqnarray}
&&\Pi _{\mu }^{\mathrm{OPE}}(p,p^{\prime })=\frac{\epsilon _{\mu \theta
\alpha \beta }}{2}\int d^{4}xd^{4}ye^{ip^{\prime }x}e^{iqy}g_{s}\frac{{%
\lambda }_{ab}^{n}}{2}G_{\alpha \beta }^{n}(0)  \notag \\
&&\times \mathrm{Tr}\left[ S_{b}^{ai}(-y)\gamma _{5}S_{d}^{ij}(y-x)\gamma
_{5}S_{b}^{jb}(x)\gamma _{5}\gamma _{\theta }\right] .  \label{eq:QCDside2}
\end{eqnarray}

The sum rule for the form factor $g_{7}(q^{2})$ is
\begin{eqnarray}
&&g_{7}(q^{2})=\frac{2m_{H_{\mathrm{b}}}m_{b}^{2}(q^{2}-m_{B}^{2})}{f_{H_{%
\mathrm{b}}}f_{B}^{2}m_{B}^{4}(m_{H_{\mathrm{b}}}^{2}+m_{B}^{2}-q^{2})}
\notag \\
&&\times e^{m_{H_{\mathrm{b}}}^{2}/M_{1}^{2}}e^{m_{B}^{2}/M_{2}^{2}}\Pi (%
\mathbf{M}^{2},\mathbf{s}_{0},q^{2}).  \label{eq:SRCoup4}
\end{eqnarray}

Eq.\ (\ref{eq:SRCoup4}) contains the mass $m_{H_{\mathrm{b}}}$ and current
coupling  $f_{H_{\mathrm{b}}}$ of the vector bottom hybrid meson $H_{\mathrm{%
b}}$. These quantities were found in Ref.\ \cite{Alaakol:2024zyh}.
\begin{eqnarray}
m_{H_{\mathrm{b}}} &=&(10.41\pm 0.18)~\mathrm{GeV},  \notag \\
f_{H_{\mathrm{b}}} &=&(12\pm 3)\times 10^{-2}~\mathrm{GeV}^{3}.  \label{eq:Result2}
\end{eqnarray}%

In numerical computations, we choose the parameters $(M_{1}^{2},s_{0})$ and $%
(M_{2}^{2},s_{0}^{\prime })$ in the following manner: We use the working
regions $M_{1}^{2}\in \lbrack 12,14]~\mathrm{GeV}^{2}$ and $s_{0}\in \lbrack
120,125]~\mathrm{GeV}^{2}$  for the $H_{\mathrm{b}}$ channel  \cite%
{Alaakol:2024zyh}, whereas for the $B^{+}$ meson channel employ
\begin{equation}
M_{2}^{2}\in \lbrack 5.5,6.5]~\mathrm{GeV}^{2},\ s_{0}^{\prime }\in \lbrack
33.5,34.5]~\mathrm{GeV}^{2}.
\end{equation}

\begin{figure}[h]
\includegraphics[width=8.5cm]{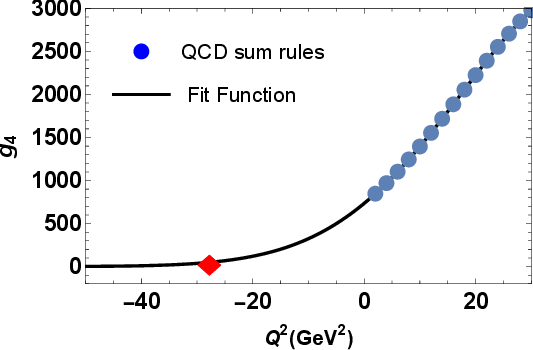}
\caption{QCD data and extrapolating function $\mathcal{F}_{4}(Q^{2})$. The
red diamond fixes the point $Q^{2}=-m_{B}^{2}$. }
\label{fig:Fit2}
\end{figure}

The strong coupling $g_{7} $ is determined at the mass shell $%
q^{2}=m_{B}^{2} $ of the $B^{-}$ meson
\begin{equation}
{g}_{7}\equiv \mathcal{F}_{7}(-m_{B}^{2})=48.71\pm 6.33.  \label{eq:Coupl4}
\end{equation}%
The function $\mathcal{F}_{7}(Q^{2})$ is fixed by the coefficients $\mathcal{%
F}_{7}^{0}=734.57$, $c_{7}^{1}=8.21$, and $c_{7}^{2}=-11.10$. The relevant
SR predictions for the form factor ${g}_{7}(Q^{2})$ and $\mathcal{F}%
_{7}(Q^{2})$ are plotted in Fig.\ \ref{fig:Fit2}. The width of the decay $H_{%
\mathrm{b}}\rightarrow B^{+}B^{-} $ is

\begin{equation}
\Gamma _{7}\left[ H_{\mathrm{b}}\rightarrow B^{+}B^{-}\right] =(39.4\pm
10.9)~\mathrm{MeV}.  \label{eq:DW4}
\end{equation}

The decay parameters of the process $H_{\mathrm{b}}\rightarrow B_{0}%
\overline{B}_{0}$ are found to be nearly identical to those of the channel $%
H_{\mathrm{b}}\rightarrow B^{+}B^{-}$ , despite the slight mass difference
between the neutral and charged $B$ mesons, with $m_{B_{0}}=(5279.72\pm
0.08)~\mathrm{MeV}$ and $m_{B^{\pm }}$ differing marginally. Consequently,
the form factors satisfy $g_{8}(q^{2})\approx g_{7}(q^{2})$, leading to an
approximate equality in decay widths: $\Gamma _{8}\left[ H_{\mathrm{b}%
}\rightarrow B_{0}\overline{B}_{0}\right] \approx \Gamma _{7}\left[ H_{%
\mathrm{b}}\rightarrow B^{+}B^{-}\right] $.

The decay processes examined in this section allow for the evaluation of the
full width of the exotic meson $H_{\mathrm{b}}$, resulting in
\begin{equation}
\Gamma _{H_{\mathrm{b}}}=(78.8\pm 15.4)~\mathrm{MeV}.  \label{eq:fullDW2}
\end{equation}

%%%%%%%%%%%%%%%%%%%%%%%%%%%%%%%%%%%%%%%%%%%%%%%%%%%%%%%%%%%%%%%%%%%%%%%%`

\section{Summing up}

\label{sec:Dis} %%%%%%%%%%%%%%%%%%%%%%%%%%%%%%%%%%%%%%%%%%%%%%%%%%%%%%%%%%%
In this work, we have performed a comprehensive analysis of the strong
decays of the vector charmonium and bottomonium hybrid mesons $H_{\mathrm{c}}
$ and $H_{\mathrm{b}}$, which possess the quantum numbers $1^{\mathrm{--}}$,
within the framework of QCD three-point sum rule method. Hybrid mesons,
characterized by explicit gluonic excitations in addition to the
quark-antiquark content, are predicted by QCD but remain among the least
understood hadronic states. Identifying their distinctive features and decay
characteristics is essential for advancing our understanding of
nonperturbative QCD dynamics and the full spectrum of hadronic matter.

We focused on the dominant decay modes of $H_{\mathrm{c}}$ and $H_{\mathrm{b}%
}$ into open-charm and open-bottom meson pairs, specifically $D^{+}D^{-}$, $%
D_{0}\overline{D}_{0}$, $D_{s}\overline{D}_{s}$, $ D^{\ast +}D^{\ast -} $, $%
D^{\ast 0}\overline{D}^{\ast 0}$, $ D^{\ast +}D^{-} $, $ D^{\ast 0}\overline{D}^{0} $, $D_{s}^{\ast +}D_{s}^{-}  $ and $B^{+}B^{-}$, $B_{0}%
\overline{B}_{0}$. By calculating the strong coupling constants at the
relevant hybrid-meson-meson interaction vertices, we evaluated the
corresponding partial decay widths and determined the full decay widths to
be $\Gamma _{H_{\mathrm{c}}}=(309.6\pm 39.0)~\mathrm{MeV}$ and $\Gamma _{H_{%
\mathrm{b}}}=(78.8\pm 15.4)~\mathrm{MeV}$. These results suggest that $H_{%
\mathrm{c}}$  may appear as a broad resonance in invariant mass
distributions, whereas $H_{\mathrm{b}}$ is relatively narrow state. These
parameters of the heavy hybrid quarkonia are promising experimental
observables for future studies in the charm and bottom sectors.

The methods and results presented provide valuable theoretical input for the
identification and classification of exotic charmonium- and bottomonium-like
hybrid states. They also contribute to testing nonperturbative QCD
predictions concerning the hybrid meson spectrum.

\section*{ACKNOWLEDGMENTS}

The author is grateful to S. S. Agaev, K. Azizi and H. Sundu for helpful
discussions.

\appendix*

\begin{widetext}

\section{ Some expressions in the OPE side of the calculations for  the decay $H_{\mathrm{c}%
}\rightarrow D^{+}D^{-}$}

\renewcommand{\theequation}{\Alph{section}.\arabic{equation}} \label{sec:App}
%%%%%%%%%%%%%%%%%%%%%%%%%%%%%%%%%%%%%%%%%%%%%%%%%%%%%%%%%%%%%%%%%%%%%%

This Appendix contains expressions of correlation functions. The correlation function $\Pi 
(\mathbf{M}^{2},\mathbf{s}_{0},q^{2})$ has the following form as presented in Eq.\ 
(\ref{eq:SCoupl}):
\begin{eqnarray}
&&\Pi (\mathbf{M}^{2},\mathbf{s}_{0},q^{2})=\int_{4m_{c}^{2}}^{s_{0}}ds%
\int_{m_{c}^{2}}^{s_{0}^{\prime }}ds^{\prime }\rho (s,s^{\prime },q^{2})
\times e^{-s/M_{1}^{2}}e^{-s^{\prime }/M_{2}^{2}}+\Pi (\mathbf{M}^{2}),  \label{eq:A1}
\end{eqnarray}
where,
\begin{equation}
\rho (s,s^{\prime },q^{2})=\int_{0}^{1}d\alpha \int_{0}^{1-\alpha }d\beta \int_{0}^{1-\alpha
-\beta }d\gamma \rho (s,s^{\prime },q^{2},\alpha ,\beta , \gamma),  \label{eq:A2}
\end{equation}
and
\begin{equation}
\Pi (\mathbf{M}^{2})=\int_{0}^{1}d\alpha \int_{0}^{1-\alpha }d\beta \int_{0}^{1-\alpha
-\beta }d\gamma \Pi (\mathbf{M}^{2},q^{2},\alpha ,\beta , \gamma).  \label{eq:A3}
\end{equation}
Here, $\alpha $, $\beta $, and $\gamma $ are the Feynman parameters.

The spectral density is found as
\begin{eqnarray}
&&\rho (s,s^{\prime },q^{2},\alpha ,\beta , \gamma) = 
\frac{g_{s}^{2} \beta \Theta ({N_{1})}}{768 \pi^{4} L_{1}^{2}}
\bigg[  
    2 m_{c} m_{d} (-3 + 2 \beta) (-1 + 2 \alpha)  \nonumber \\
&& + \ m_{c}^{2} \big(
        \beta^{2} (3 - 6 \alpha) 
        + 2 \alpha (1 + \alpha) 
        + \beta (2 + 5 \alpha - 6 \alpha^{2})
      \big) \nonumber \\
&& + \ s^{\prime} \beta \big(
        \beta^{2} (4 - 8 \alpha) 
        + \beta (13 - 7 \alpha) \alpha 
        + 2 (-2 + \alpha + \alpha^{2})
      \big) \nonumber \\
&& - \ \alpha \Big(
        s \beta (2 + 4 \beta + 2 \alpha - 7 \beta \alpha) 
        + q^{2} \big(
            2 - 2 \alpha^{2} 
            + \beta^{2} (-4 + 7 \alpha) 
            + \beta (1 - 11 \alpha + 6 \alpha^{2})
          \big)
      \Big)
\bigg] \notag \\
&& +
\frac{g_s^2 \alpha \Theta ({N_{2})}}{2048 \pi^4 D^6} 
\Big[
-6 s^{\prime }\beta A_1(\alpha,\beta) - 2 \gamma^6 (3 s^{\prime} \beta + 2 q^2 \alpha) \notag \\
&& + 2 m_c m_d (-1 + \gamma + 2 \alpha) D^2 D_1 - 2 \gamma^5 A_2 + \gamma \alpha^2 A_3 \notag \\
&& - 2 \gamma^4 A_4 + m_c^2 D A_5 - \gamma^3 A_6 + \gamma^2 \alpha A_7
\Big] \notag \\
&& +\frac{g_s^2 L_2\Theta ({N_{2})}}{64 \pi^4 (-1 + \gamma) D^5} \Bigg\{ m_c^2 \big[ 3 \gamma^5 (\beta + \alpha) + 3 \gamma^4 (2 \beta^2 + \alpha (-3 + 2 \alpha) \notag \\
&& + \beta (-3 + 4 \alpha)) + \gamma^3 B_1 + \gamma^2 B_2 + \gamma B_3 + B_4 \big] \notag \\
&& - q^2 \beta  \big[ 3 \gamma^7 + \gamma^6 (-12 + 9 \beta + 13 \alpha) + \gamma^5 (18 + 6 \beta^2 - 48 \alpha + 29 \alpha^2 + \beta (-27 + 43 \alpha)) \notag \\
&&  + \alpha^3 (-9 \beta^3 + \beta (-18 + \alpha)(-1 + \alpha)^2 - 3 (-1 + \alpha)^3 + \beta^2 (24 - 25 \alpha + \alpha^2)) + \gamma^4 B_5 \notag \\
&&+ \gamma^3 B_6 + \gamma^2 B_7 + \gamma \alpha^2 B_8 \big] - \gamma \alpha  \big[ 4 \gamma^6 s^{\prime } + \gamma^5 (-4 s \beta + s^{\prime } (-16 + 11 \beta + 12 \alpha)) \notag \\
&&+ \gamma^4 (s \beta (12 - 7 \beta - 11 \alpha) + s^{\prime } C_1) + \alpha^2 (s \beta C_2 + s^{\prime } C_3) + \gamma^3 (s \beta C_4 + s^{\prime } C_5) \notag \\
&&+ \gamma \alpha (s \beta C_6 + s^{\prime } C_7) + \gamma^2 (- s \beta C_8 + s^{\prime } C_9) \big] \Bigg\}, 
\label{eq:A2}
\end{eqnarray}

where
\begin{eqnarray}
&&D = \gamma^2 + \gamma(-1+\alpha) + (-1+\alpha) \alpha ,\ \notag \\
&&D_1 = \gamma^2 + \gamma(-1 + \alpha) + \alpha(-1 + 6 \beta + \alpha),\   \notag \\
&&A_1 = \beta (6 \beta^2 + 7 \beta (-1 + \alpha) + (-1 + \alpha)^2) (-1 + \alpha) \alpha^3 , \notag \\
&&A_2 = -3 s \beta \alpha + q^2 \alpha (-6 + 3 \beta + 10 \alpha) + s^{\prime}\beta (-9 + 3 \beta + 17 \alpha),\notag \\
&&A_3 = 2 \bigl( s \beta (-3 \beta^2 + 19 \beta (-1+\alpha) + 3 (-1+\alpha)^2) + q^2 (3 \beta^3 - 15 \beta^2 (-1+\alpha) \notag \\ 
&&- 15 \beta (-1+\alpha)^2 - 2 (-1+\alpha)^3) \bigr) \alpha
+ s^{\prime} \beta \bigl( \beta^2 (72 - 137 \alpha) - 2 (-1+\alpha)^2 (-9 + 17 \alpha) \notag \\
&&+ \beta (-90 + 281 \alpha - 191 \alpha^2) \bigr),\notag \\
&&A_4 = s^{\prime} \beta (9 - 43 \alpha + 39 \alpha^2 + \beta (-6 + 33 \alpha)) + \alpha \bigl( s \beta (6 - 11 \alpha) + q^2 (6 - 6 \beta - 24 \alpha + 23 \beta \alpha + 18 \alpha^2) \bigr),\notag \\
&&A_5= 4 \gamma^4 (\beta + \alpha) + 8 \gamma^3 (\beta + \alpha)(-1 + 2 \alpha) + 4 \gamma^2 (\beta + \alpha - 6 \beta \alpha + 6 \beta^2 \alpha - 6 \alpha^2 + 11 \beta \alpha^2 + 5 \alpha^3) \notag \\
&&+ 4 (-1 + \alpha) \alpha^2 (6 \beta^2 + (-1 + \alpha) \alpha + \beta (-1 + 7 \alpha)) + \gamma \alpha (3 \beta^2 (-8 + 21 \alpha) + 8 \alpha (1 - 3 \alpha + 2 \alpha^2) \notag \\
&&+ \beta (8 - 48 \alpha + 79 \alpha^2)),\notag \\
&&A_6= \alpha \bigl(-2 s \beta (3 - 17 \alpha + 19 \beta \alpha + 14 \alpha^2) + q^2 (-4 + 6 \beta + 36 \alpha - 82 \beta \alpha + 30 \beta^2 \alpha - 68 \alpha^2 + 115 \beta \alpha^2 + 36 \alpha^3) \bigr) \notag \\
&&+ s^{\prime} \beta \bigl( 36 \beta^2 \alpha + \beta (6 - 120 \alpha + 215 \alpha^2) + 2 (-3 + 35 \alpha - 82 \alpha^2 + 50 \alpha^3) \bigr), \notag \\
&&A_7 = s^{\prime} \beta \bigl( \beta^2 (36 - 137 \alpha) + \beta (-54 + 305 \alpha - 304 \alpha^2) + 2 (9 - 52 \alpha + 82 \alpha^2 - 39 \alpha^3) \bigr)\notag \\
&&+ \alpha \bigl( 2 s \beta (6 - 3 \beta^2 - 17 \alpha + 11 \alpha^2 + \beta (-19 + 48 \alpha)) + q^2 (6 \beta^3 + \beta^2 (30 - 75 \alpha) \notag \\
&&- 4 (-1 + \alpha)^2 (-2 + 5 \alpha) + \beta (-36 + 145 \alpha - 109 \alpha^2)) \bigr), \notag \\
&&B_1 = 9 (-1 + \alpha)^2 \alpha + 3 \beta^2 (-4 + 7 \alpha) + \beta (9 - 30 \alpha + 31 \alpha^2), \notag \\
&&B_2 =9 \beta^3 \alpha + 6 \beta^2 (1 - 7 \alpha + 6 \alpha^2) + 3 \alpha (-1 + 6 \alpha - 7 \alpha^2 + 2 \alpha^3) + \beta (-3 + 24 \alpha - 65 \alpha^2 + 34 \alpha^3), \notag \\
&&B_3 = 9 \beta^3 (-1 + \alpha) + 3 (-2 + \alpha) (-1 + \alpha)^2 \alpha + \beta^2 (21 - 51 \alpha + 25 \alpha^2) + \beta (-6 + 37 \alpha - 56 \alpha^2 + 20 \alpha^3),  \notag \\
&&B_4 = \alpha^2 \big(-9 \beta^3 - 3 (-1 + \alpha)^2 \alpha + \beta^2 (15 - 25 \alpha + \alpha^2) + \beta (-3 + 22 \alpha - 20 \alpha^2 + \alpha^3) \big), \notag \\
&&B_5 = -12 + 66 \alpha - 96 \alpha^2 + 39 \alpha^3 + 12 \beta^2 (-1 + 3 \alpha) + \beta (27 - 122 \alpha + 100 \alpha^2),  \notag \\
&& B_6 = 3 - 40 \alpha + 9 \beta^3 \alpha + 114 \alpha^2 - 109 \alpha^3 + 32 \alpha^4 + \beta^2 (6 - 72 \alpha + 80 \alpha^2) + \beta (-9 + 115 \alpha - 245 \alpha^2 + 125 \alpha^3), \notag \\
&&B_7 = 9 - 56 \alpha + 104 \alpha^2 - 73 \alpha^3 + 16 \alpha^4 + 9 \beta^3 (-1 + 2 \alpha) + \beta^2 (36 - 134 \alpha + 75 \alpha^2) \notag \\
 &&+ \beta (-36 + 190 \alpha - 246 \alpha^2 + 79 \alpha^3), \notag \\
&& B_8 = 9 \beta^3 (-2 + \alpha) + (-1 + \alpha)^2 (9 - 19 \alpha + 3 \alpha^2) + \beta^2 (54 - 99 \alpha + 26 \alpha^2) \notag \\
&&+ \beta (-45 + 139 \alpha - 116 \alpha^2 + 22 \alpha^3),\notag \\
 &&C_1 = 24 - 33 \beta + 7 \beta^2 - 44 \alpha + 45 \beta \alpha + 20 \alpha^2 , \notag \\
&&C_2 = 15 \beta^2 + 7 (-1 + \alpha)^2 + \beta (-25 + 27 \alpha - 2 \alpha^2) ,\notag \\
 &&C_3 = -12 \beta^3 - 4 (-1 + \alpha)^3 + \beta (-1 + \alpha)^2 (-26 + 3 \alpha) + 2 \beta^2 (17 - 18 \alpha + \alpha^2) ,\notag \\
 &&C_4 = -12 + 14 \beta + 33 \alpha - 32 \beta \alpha - 18 \alpha^2 ,\notag \\
 &&C_5 = \beta^2 (-14 + 41 \alpha) + \beta (33 - 127 \alpha + 88 \alpha^2) + 4 (-4 + 15 \alpha - 16 \alpha^2 + 5 \alpha^3), \notag \\
 &&C_6 = 11 - 15 \beta^2 (-1 + \alpha) - 32 \alpha + 28 \alpha^2 - 7 \alpha^3 + \beta (-32 + 72 \alpha - 27 \alpha^2),\notag \\
 &&C_7 = 12 \beta^3 (-1 + \alpha) + 4 (-1 + \alpha)^2 (2 - 4 \alpha + \alpha^2) + \beta^2 (41 - 100 \alpha + 36 \alpha^2) + \beta (-37 + 140 \alpha - 135 \alpha^2 + 32 \alpha^3), \notag \\
 &&C_8 = -4 + 33 \alpha + 15 \beta^2 \alpha - 43 \alpha^2 + 14 \alpha^3 + \beta (7 - 64 \alpha + 47 \alpha^2),  \notag \\
 &&C_9 = 12 \beta^3 \alpha + 4 (-1 + \alpha)^2 (1 - 7 \alpha + 3 \alpha^2) + \beta^2 (7 - 82 \alpha + 66 \alpha^2) + \beta (-11 + 119 \alpha - 202 \alpha^2 + 80 \alpha^3).
\end{eqnarray}

In Eq.\ (\ref{eq:A2}), $\Theta ({N)}$ is the Unit Step function. Here

\begin{eqnarray}
 &&N_1 = - s^{\prime} \, \beta L_{1} - m_{c}^{2} (\beta + \alpha) + \alpha \left( s \beta - q^{2} L_{1} \right)
, \notag \\
 &&N_2 = - \frac{(\gamma - 1) \left[ -s^{\prime}\,\beta\,(\gamma + \alpha)L_{2} 
+ m_c^2 (\beta + \alpha) \left( \gamma^2 + \gamma(\alpha - 1) + \alpha(\alpha - 1) \right) 
+ \gamma \alpha \left( s\beta - q^{2} L_{2} \right) \right]}
{\left[ \gamma^2 + \gamma(\alpha - 1) + \alpha(\alpha - 1) \right]^2}.
\end{eqnarray}

We also use the notations

\begin{equation}
L_{1}=\alpha +\beta -1,\ L_{2}=\alpha +\beta +\gamma -1.
\end{equation}

Components of the function $ \Pi (\mathbf{M}^{2}) $ are:

\begin{eqnarray}
&& \Pi ^{\mathrm{Dim3}}(\mathbf{M}^{2},q^{2},\alpha ,\beta ) =
\frac{ g_s^{2}}{96 \pi^{2}M_{2}^{2}  \beta  L_{1}^2} 
\langle \overline{d} d \rangle 
\exp\left[
  \frac{q^{2} \alpha L_{1} + m_{c}^2 (L_{1}+1)}
       {M_{2}^{2}  \beta  L_{1}}
\right]  \notag \\
&&\times\Bigg[\big( 2 m_{c} M_{2}^{2} \beta (1 - 2 \alpha) 
      + m_{d} (q^2 - M_{2}^{2} \beta) \alpha L_{1} 
      + m_{c}^2 m_{d} (L_{1}+1) \big) 
\delta\Big(\frac{1}{M_{1}^{2}} - \frac{\alpha}{M_{2}^{2} L_{1}}\Big) \notag \\
&& - m_{d}\beta \alpha  \delta^{\prime}\Big(\frac{1}{M_{1}^{2}} - \frac{\alpha}{M_{2}^{2} L_{1}}\Big)
\Bigg],
\end{eqnarray}

\begin{eqnarray}
&&\Pi ^{\mathrm{Dim4}}(\mathbf{M}^{2},q^{2},\alpha ,\beta ,\gamma) =\langle \alpha_s G^2 / \pi \rangle \exp\Bigg[\frac{q^2 \alpha L_1 + m_c^2 (L_1+1)}{M_2^2 \beta L_1}\Bigg] \Bigg\{
\frac{ g_s^2 m_c^2}
{2304 M_2^4 \pi^2 \gamma^3 \beta \alpha^2 L_1 L_2^2 D^3} \notag \\
&& \times  
\Bigg[
(\gamma^2 + \gamma(-1+\alpha) + (-1+\alpha)\alpha) \Bigg(
2 m_c^2 \gamma (\beta+\alpha)^2 (\beta^2 - \beta \alpha + \alpha^2) \notag \\
&& \times
\Big(
\gamma^3 + 2 \gamma^2 (\beta+\alpha) + \alpha (\beta+\alpha)^2 
+ \gamma((-1+\beta)^2 + (-3+4\beta)\alpha + 2 \alpha^2) \notag \\
&&  - (\beta+\alpha)(\gamma^2 (-1+\beta) + \gamma \alpha (3-4\beta+2\alpha) + \alpha^2 (1-2\beta+\alpha))
\Big) \delta\Big(\frac{1}{M_1^2} - \frac{\alpha}{M_2^2 L_1}\Big) \notag \\
&&  - 2 \gamma \beta (\beta^3 + \alpha^3) 
\delta'\Big(\frac{1}{M_1^2} - \frac{\alpha}{M_2^2 L_1}\Big)
\Bigg)\Bigg] + \frac{1}{96 M_2^2 \beta^2 L_1^3} 
\Bigg[
- \Big(
3 m_c m_d L_1 (\beta+\alpha) (1 - 2 \alpha) \notag \\
&& + m_c^2 (\beta+\alpha) (1 - \beta + \beta (1 - 2 \beta) \alpha + (-2 + 3 \beta) \alpha^2 + \alpha^3) - L_1 (- q^2 \alpha (1 + (\beta+\alpha) (\beta + (-2 + \beta) \alpha + \alpha^2)) \notag \\
&& + M_2^2 \beta (2 + \beta^2 \alpha + 2 \alpha (-3 + \alpha + \alpha^2) + \beta (-2 + \alpha + 3 \alpha^2)))) \delta\Big(\frac{1}{M_1^2} - \frac{\alpha}{M_2^2 L_1}\Big) 
\Big) \notag \\
&&+ \beta \alpha ((-1 + \beta) \beta + \alpha + \beta (1 + \beta) \alpha + (-1 + \beta) \alpha^2) 
\delta'\Big(\frac{1}{M_1^2} - \frac{\alpha}{M_2^2 L_1}\Big)
\Bigg]
\Bigg\},
\end{eqnarray}

where $ \delta' = \frac{\partial \delta}{\partial(\frac{1}{M_1^2})} $.

\end{widetext}

\renewcommand{\theequation}{\Alph{section}.\arabic{equation}} \label{sec:App}

%%%%%%%%%%%%%%%%%%%%%%%%%%%%%%%%%%%%%%%%%%%%%%%%%%%%%%%%%%%%%%%%%%%%%%%%%%%%%%%%%%
%%%%%%%%%%%%%%%%%%%%%%%%%%%%%%%%%%%%%%%%%%%%%%%%%%%%%%%%%%%%%%%%%%%%%%%%%%%%%%%%%%

\end{document}